\documentclass{elsart}

\usepackage{natbib}
\bibpunct{(}{)}{;}{a}{}{,}

 \usepackage{graphics}
 \usepackage{graphicx}
\usepackage{epsfig}

\usepackage{amssymb}

\def\Rate{{\it R}}

\begin{document}

\newcommand {\peryr}{$\rm{yr}^{-1}$}
\newcommand {\permyr}{$\rm{Myr}^{-1}$}
\newcommand{\nsns}{NS$-$NS\ }
\newcommand{\nswd}{NS$-$WD\ }
\newcommand{\wdns}{WD$-$NS\ }
\newcommand{\tlife}{\tau_{\rm life}}
\newcommand{\tmrg}{$\tau_{\rm mrg}$}
\newcommand{\td}{$\tau_{\rm d}$}
\newcommand{\tc}{$\tau_{\rm c}$}
\newcommand{\tsd}{$\tau_{\rm sd}$}
\newcommand{\rate}{$\cal R$}
\newcommand{\rtot}{${\cal R}_{\rm tot}$}
\newcommand{\rpeak}{${\cal R}_{\rm peak}$}
\newcommand{\rdet}{${\cal R}_{\rm det}$}
\newcommand{\ntot}{$N_{\rm tot}$}
\newcommand{\npsr}{N_{\rm PSR}}
\newcommand{\nobs}{$N_{\rm obs}$}
\newcommand{\nmean}{$<N_{\rm obs}>$}
\newcommand{\solarM}{M$_{\rm \odot}$}
\newcommand{\chmass}{$\cal M$}
\newcommand{\fmin}{$f_{\rm min}$}
\newcommand{\fmax}{$f_{\rm max}$}

\begin{frontmatter}



\title{Formation of Double Compact Objects}


\author[1]{V.\ Kalogera},\author[2]{ K.\ Belczynski}, \author[3]{ C.\ Kim}, \author[1]{ R.\ O'Shaughnessy},\\ \hspace{0.2in} \author[1]{ \& B.\ Willems}

\address[1]{Department of Physics \& Astronomy, Northwestern University, 2145 Sheridan Rd, Evanston, IL 60208}

\address[2]{Department of Astronomy, New Mexico State University, 1320 Frenger Mall, Las Cruces, NM 88003}

\address[3]{Department of Astronomy, Cornell University, 522 Space Sicences Building, Ithaca, NY 14853} 

\begin{abstract}

Current observations of double neutron stars provide us with a wealth of information that we can use to investigate their evolutionary history and the physical conditions of neutron star formation. 
Understanding this history and formation conditions further allow us to make theoretical predictions for the formation of other double compact objects with one or two black hole components and assess the detectability of such systems by ground-based gravitational-wave interferometers. In this paper we summarize our group's body of work in the past few years and we place our conclusions and current understanding in the framework of other work in this area of astrophysical research. 

\end{abstract}


\end{frontmatter}

\section{Introduction}
\label{Intro}

Gravitational waves were theoretically predicted many decades ago by Einstein and his theory of General Relativity, and their existence has been indirectly confirmed by the pioneering observations of the Hulse-Taylor binary pulsar \citep{1975ApJ...195L..51H}. In the coming years, the physics and astronomy community anticipates the first direct detection of gravitational waves (GW) and the birth of a new research field: GW astrophysics, where observations with the first ground-based interferometers (such as LIGO, GEO, and Virgo) contribute to the astrophysical understanding of the sources. Double neutron stars (DNS) and their inspiral due to GW emission represent the primary target source. This is primarily because they are known to exist in the Galaxy, with the Hulse-Taylor binary and the binary pulsar \citep{2003Natur.426..531B} being the staples of this class of binary compact objects. Double compact objects with black hole (BH) components (i.e., BH-NS and BH-BH) have not yet been discovered in nature, but our current understanding  of binary star evolution naturally predicts such BH binaries that would coalesce within a Hubble time \citep{2002ApJ...572..407B, 1995ApJ...440..270B, 1998AA...332..173P, 2005ApJ...633.1076O, 2002MNRAS.329..897H, 1996AA...309..179P, 1998ApJ...496..333F, 1999ApJ...517..318B}. 

The sample of known DNS in the Galaxy has been very well studied and monitored since their discovery and as a result there is a wealth of information about their physical properties. This observational sample provides us with a unique opportunity to investigate the evolutionary history of these systems and a number of studies have addressed this basic question from different points of view.  Another way to address the question of the formation of double compact objects is to develop population synthesis models that follow the evolution of large ensembles of binaries. Such models allow the detailed study of evolutionary channels that lead to double compact object formation and assess their relative efficacy and the main binary evolution phases that affect the final properties of the binaries. 

The calculation of inspiral rates for double compact objects and associated event rates for ground-based GW detectors lies at the center of research activity related to GW searches in the last decade. These calculations were pioneered by \citet{1991ApJ...380L..17P} and \citet{1991ApJ...379L..17N}. Since then, a number of studies have considered this problem and revised these initial results, following the methods introduced in 1991. A few years ago \citet{2003ApJ...584..985K} introduced a novel statistical method for the calculation of inspiral rates based on the observed systems (empirical estimates); this method allowed us for the first time to associate rate estimates with statistical and systematic uncertainties and to place these estimates on the basis of confidence levels. Inspiral rates have also been computed based on population synthesis calculations that explore a limited number of model assumptions. More recently \citet{2005ApJ...633.1076O} have combined the two methods and have used the empirical DNS estimates as constraints on model predictions, especially for BH binaries on which we have no direct guidance from observations.

In this paper we summarize our group's work in this area in the context of double compact object formation, their properties and inspiral rates and in connection to other studies that have appeared in the literature.

\section{The Recent Evolutionary History of Known Close Double Neutron Stars}
\label{recent}

According to our current
knowledge \citep[for more details see][]{1991PhR...203....1B,
1995ApJ...440..270B, 2002ApJ...572..407B, 2003astro.ph..3456T}, double neutron stars (DNSs) form from primordial binaries in which, possibly after some initial
mass-transfer phase, both component stars have masses in excess of
$\sim 8-12\,M_\odot$. After the primary explodes in a supernova (SN)
explosion to form the first neutron star (NS), the binary enters a
high-mass X-ray binary phase in which the NS accretes matter from the
wind of its companion. The phase ends when the companion star evolves
of the main sequence and engulfs the NS in its expanding envelope. The
NS then spirals in towards the core of the companion until enough
orbital energy is transferred to the envelope to expel it from the
system. When the envelope is ejected, the binary consists of the NS
and the stripped-down helium core of its former giant companion,
orbiting each other in a tight orbit. If the NS is able to accrete
during its rapid inspiral, a first ``recycling'' may take place during
which it is spun up to millisecond periods. The helium star then
evolves further until it, in turn, explodes and forms a NS. Depending
on the mass of the helium star and the size of the orbit at the time
of the explosion, the SN may be preceded by a second recycling phase
when the helium star fills its Roche lobe and transfers mass to the
NS \citep{2001ApJ...550L.183B, 2002ApJ...571L.147B, 2002ApJ...572..407B}.

In what follows in this section we use this basic evolutionary framework and the complete, up-to-date set of observational constraints (at the time of this writing) on three DNS Galactic systems to investigate the physical properties of progenitor binaries at the time of the second SN explosion. We primarily summarize the work presented in \citet{2006PhRvD..74d3003W} and \citet{2004ApJ...616..414W} with many comparisons to other published studies addressing similar questions.

\subsection{PSR~J0737-3039: The Double Pulsar} 

The recent discovery of the strongly relativistic binary pulsar
\citep{2003Natur.426..531B} which is also the first eclipsing double pulsar
system found in our Galaxy \citep{2004Sci...303.1153L} has resparked the
interest in the evolutionary history and formation of DNS systems \citep{2004ApJ...603L.101W, 2004MNRAS.349..169D, 2004ApJ...616..414W}. 
Shortly after the discovery of PSR\,J0737-3039,
\citet{2004MNRAS.349..169D} and \citet{2004ApJ...603L.101W} independently derived that, right before the second SN
explosion, the binary was so tight that the helium star must have been
overflowing its critical Roche lobe. This conclusion, for the first
time, strongly confirmed the above formation channel for DNS
binaries. The observed 22.7\,ms pulsar (hereafter PSR\,J0737-3039A or
pulsar~A) then corresponds to the first-born NS, and its 2.8\,s pulsar
companion (hereafter PSR\,J0737-3039B or pulsar~B) to the second-born
NS.

The main properties of the double pulsar system and its component stars
relevant to this investigation are summarized in Table~1 of \citet{2006PhRvD..74d3003W}. 

The kinematic properties of PSR\,J0737-3039 have undergone significant
revision since \citet{2004ApJ...609L..71R} used interstellar
scintillation measurements and inferred a velocity component
perpendicular to the line-of-sight of $\sim 141\,{\rm
km\,s^{-1}}$. \citet{2005ApJ...623..392C} derived a reduced velocity
of $\sim 66\,{\rm km\,s^{-1}}$ by incorporating anisotropies of the
interstellar medium in the interstellar scintillation model. This
reduced velocity, however, strongly depends on the adopted anisotropy
model. \citet{2005astro.ph..3386K}, on the other hand, used pulsar
timing measurements to derive a firm model-independent upper limit of
$30\,{\rm km\,s^{-1}}$ on the transverse velocity.

The kick imparted to pulsar~B at birth is expected to tilt the orbital
plane and misalign pulsar~A's spin axis with respect to the post-SN
orbital angular momentum axis \citep[e.g.,][]{2000ApJ...541..319K}. The
degree of misalignment depends on both the magnitude and the direction
of the kick, and therefore yields a valuable piece of information on
pulsar~B's natal kick velocity. In \citet{2004ApJ...616..414W} we showed that for the pre- and post-SN orbital
parameters compatible with all available observational constraints,
misalignment angles between approximately $70^\circ$ and $110^\circ$
are highly unlikely. \citet{2005ApJ...621L..49M} derived observational
constraints on the spin-orbit misalignment based on the stability of
pulsar~A's mean pulse profile over a time span of 3 years. The authors
concluded the angle to be smaller than $\sim 60^\circ$ or larger than
$\sim 120^\circ$, in agreement with the theoretical predictions of \citet{2004ApJ...616..414W}. These misalignment limits inferred by \citet{2005ApJ...621L..49M} are incorporated in the list of available observational
constraints.

Among the constraints inferred from observations, the most uncertain
parameter affecting the reconstruction of the system's formation and
evolutionary history is its age. Characteristic ages, defined
as half the ratio of the spin period to the spin-down rate, are the
most commonly adopted age estimators, but are known to be quite
unreliable \citep[e.g.][]{2003ApJ...593L..31K}. In the case of
PSR\,J0737-3039, the spin evolution of pulsar~B is furthermore very
likely affected by torques exerted by pulsar~A on pulsar~B \citep{2004MNRAS.353.1095L}, adding to
the uncertainties of its age. \citet{2005ASPC..328..113L} therefore
derived an alternative age estimate by noting that, according to the
standard DNS formation channel, the time expired since the end of
pulsar~A's spin-up phase should be equal to the time expired since the
birth of pulsar~B. By combining this property with a selection of
different spin-down models, the authors derived a {\em most likely}
age of 30-70\,Myr, but were unable to firmly exclude younger and older
ages. A third age estimate can be obtained by assuming that pulsar~A
was recycled to the maximum spin rate and calculating the time
required for it to spin down to the currently observed value. This
so-called spin-down age provides an upper limit to the age of the
system of 100\,Myr \citep{2003Natur.426..531B}. In view of these
uncertainties, we derive constraints on the formation and evolution of
PSR\,J0737-3039 for three different sets of age assumptions: (i) $0
\le \tau \le 100$\,Myr,
(ii) $30 \le \tau \le 70$\,Myr (the most likely age range derived by
\citet{2005ASPC..328..113L}), and (iii) $\tau \simeq 50$\,Myr (the
characteristic age of pulsar~B).

\subsubsection{Basic assumptions and outline of the calculation}
\label{basic}

In the following subsections we use the presently known
  observational constraints to reconstruct the evolutionary
history of PSR\,J0737-3039 and constrain its properties at the
formation time of pulsar~B. More specifically, our aim is to derive a
probability distribution function (PDF) for the magnitude and direction of the kick velocity imparted to
pulsar~B at birth, the mass of pulsar~B's progenitor
immediately before it explodes in a SN explosion, and the orbital
separation of the component stars right before the SN explosion.
Other recent studies with similar goals include \citet{2006astro.ph..3649P} and \citet{2006MNRAS.tmpL.105S}. Adopting the standard DNS formation channel, the progenitor of the
double pulsar right before the formation of pulsar~B, consists of the
first-born NS, pulsar~A, and the helium star progenitor of the
second-born NS. Since the formation of the second NS is preceded by
one or more mass-transfer phases, tidal forces can safely be assumed
to circularize the orbit prior to the formation of pulsar~B.  In what follows, we refer to the times right before and right
after the formation of pulsar~B by pre-SN and post-SN,
respectively. The analysis presented consists of four major parts.

Firstly, the motion of the system in the Galaxy is traced back in time
up to a maximum age of $100$\,Myr. The goal of this calculation is to
derive the position and center-of-mass velocity of the binary right
after the formation of pulsar~B, and use this information to constrain
the kick imparted to it at birth. In order to determine possible birth
sites, our analysis is supplemented with the assumptions that the
primordial DNS progenitor was born close to the Galactic plane, and
that the first SN explosion did not kick the binary too far out of it.
The first assumption is conform with our current knowledge that
massive stars form and live close to the Galactic plane (their typical
scale height is $\simeq 50$--70\,pc). The second assumption
on the other hand neglects a small fraction of systems formed with
high space velocities \citep{2002ApJ...574..364P, 2002ApJ...572..407B}. Under these
assumptions, the binary is still close to the Galactic plane at the
formation time of pulsar~B. Possible birth sites can thus be obtained
by calculating the motion in the Galaxy backwards in time and looking
for crossings of the orbit with the Galactic mid-plane. The times in
the past at which the plane crossings occur yield kinematic estimates
for the age of the DNS, while comparison of the system's total
center-of-mass velocity with the local Galactic rotational velocity at
the birth sites yields the system's post-SN {\em peculiar} velocity.

Secondly, the orbital semi-major axis and eccentricity right after the
SN explosion forming pulsar~B are determined by integrating the
equations governing the evolution of the orbit due to gravitational
wave emission backwards in time. The integration is performed for each
Galactic plane crossing found from the Galactic motion
calculations. Since each crossing yields a different kinematic age,
and thus a different endpoint of the reverse orbital evolution calculation,
the post-SN parameters are a function of the considered Galactic plane
crossing.

Thirdly, the conservation laws of orbital energy and angular momentum
are used to map the post-SN binary parameters to the pre-SN ones. The
mapping depends on the kick velocity imparted to pulsar~B at birth and
results in constraints on the magnitude and direction of the kick
velocity, the mass of pulsar~B's pre-SN helium star progenitor, and
the pre-SN orbital separation. The pre-SN orbital eccentricity is
assumed to be zero, as expected from strong tidal forces operating on
the binary during the mass-transfer phase(s) responsible for spinning
up pulsar~A. The kick and pre- and post-SN binary parameters are then
furthermore subjected to the requirements that they be consistent with
the post-SN peculiar velocity obtained from the Galactic motion
calculations and with the observationally inferred spin-orbit
misalignment angle of pulsar~A. The latter constraint requires the
additional assumption that tidal forces align the pre-SN rotational
angular momentum vector of pulsar~A with the pre-SN orbital angular
momentum vector.

Fourthly, the solutions of the conservation laws of orbital energy and
angular momentum are weighted according to the likelihood that they
lead to the system's currently observed position and velocity in the
Galaxy. A PDF of the admissible kick velocity and progenitor
parameters is then constructed by binning the solutions in a
multi-dimensional ``rectangular'' grid. The maximum of the resulting
PDF yields the most likely magnitude and direction of the kick
velocity imparted to pulsar~B at birth, mass of pulsar~B's pre-SN
helium star progenitor, and pre-SN orbital separation. We also
investigate the sensitivity of the PDF to the adopted assumptions by
systematically varying the underlying assumptions, such as, e.g., the
age and the magnitude of the transverse systemic velocity component.

\subsubsection{Kinematic age and history}

Tracing the motion of PSR\,J0737-3039 back in time requires the
knowledge of both its present-day position and 3-dimensional space
velocity. Unfortunately, no method is presently available to measure
radial velocities of DNSs, so that the knowledge of their space
velocity is limited to the component perpendicular to the
line-of-sight (transverse velocity). At the time of this analysis, only an upper
limit was available for the double pulsar\footnote{At present a measurement of $10\,{\rm
km\,s^{-1}}$ has been reported by \citet{2006Sci...314...97K}.}. The direction of the motion in the plane perpendicular
to the line-of-sight is still unknown. Therefore we explore the system's kinematic history in terms of two
unknown parameters: (i) the radial component $V_r$ of the present-day
systemic velocity, and (ii) the orientation $\Omega$ of the transverse
velocity in the plane perpendicular to the line-of-sight. Since the
results presented in this paper do not depend on the exact definition
of $\Omega$, we refer the reader to \citet{2004ApJ...616..414W} for a more detailed
discussion and definition. We calculate the motion of the system in the Galaxy
backwards in time using the Galactic potential of
\citet{1987AJ.....94..666C} with updated model parameters derived by
\citet{1989MNRAS.239..571K}. Since at the time of the analysis kinematical constraints provided only an upper limit of 30\,km\,$s^{-1}$ on the transverse
systemic velocity, and the calculation of the past orbit requires
precise starting values for both the present-day position and the
present-day velocity, we calculate the motion backwards in time for
two specific velocities: $V_t=10$\,km\,s$^{-1}$ and
$V_t=30$\,km\,s$^{-1}$ in what follows. 

 The number of Galactic
plane crossings associated with each $V_r$ and $\Omega$, within the
adopted age limit of 100\,Myr, can be anywhere between 1 and 5. The
times in the past at which the system crosses the Galactic plane
furthermore yield kinematic estimates for the age of the DNS, while
subtraction of the Galactic rotational velocity from the total
systemic velocity at the birth sites yields the peculiar velocity
right after the formation of pulsar~B.

 Since there are no a priori constraints on the magnitude of the
  radial velocity, we examine the effect of incorporating large
radial velocities in the analysis by weighting each considered radial
velocity according to a pre-determined radial velocity
distribution. For this purpose, we performed a population synthesis
study of Galactic DNSs, including their kinematic evolution in the
potential of \citet{1987AJ.....94..666C}. Theoretical radial velocity
distributions are then obtained by taking a snapshot of the population
at the current epoch and determining the radial velocity for each DNS
in the sample. The resulting PDFs are found to be represented well by
Gaussian distributions with means of 0\,km\,s$^{-1}$ and velocity
dispersions of 60--200\,km\,s$^{-1}$, depending on the adopted
population synthesis assumptions. For comparison, we also consider a uniform radial velocity
distribution in which all radial velocities between
-1500\,km\,s$^{-1}$ and 1500\,km\,s$^{-1}$ are equally
probable. For the other unknown parameter, the orientation angle $\Omega$ of the transverse velocity component in the plane perpendicular to the line-of-sight, we assume a uniform
distribution between $0^\circ$ and $360^\circ$.

\begin{figure*}
\resizebox{14.5cm}{!}{\includegraphics{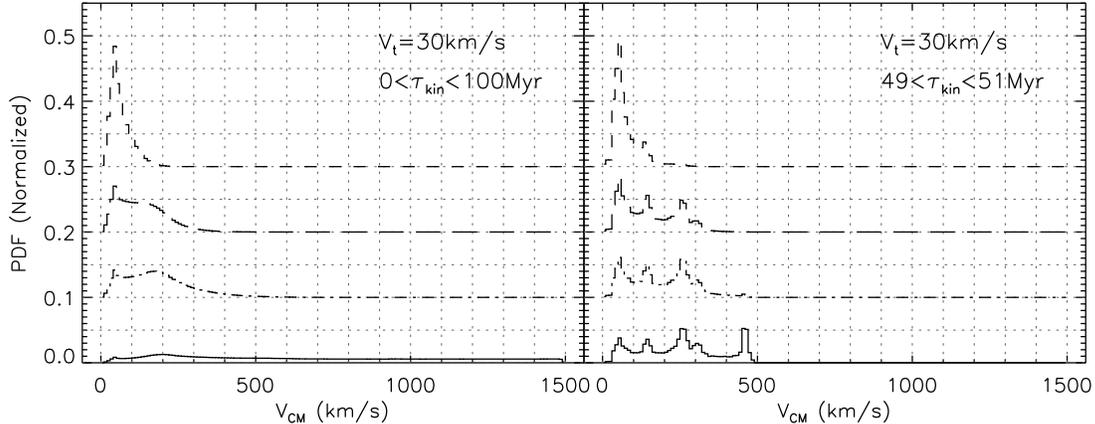}}
\caption{Distribution of post-SN peculiar velocities for a present-day
  transverse velocity of 30\,km\,s$^{-1}$, kinematic ages ranges of
  0--100\,Myr and 49--51\,Myr, and radial velocity distributions
  varying from a uniform distribution (solid line) to Gaussian
  distributions with velocity dispersions of 60\,km\,s$^{-1}$
  (long-dashed line), 130\,km\,s$^{-1}$ (short-dashed line), and
  200\,km\,s$^{-1}$ (dash-dotted line). For clarity, the PDFs are
  offset from each other by an arbitrary amount.}
\label{vcm}
\end{figure*}
 
 In Fig.~\ref{vcm}, the distribution of
post-SN peculiar velocities obtained by tracing the motion of
  PSR\,J0737-3039 in the Galaxy back in time is shown for a present-day transverse
velocity component of 30\,km\,s$^{-1}$, and kinematic age ranges of
0-100\,Myr and 49-51\,Myr. These probability distributions are
calculated considering weights according to (i) the probability that
the system has a present-day radial velocity $V_r$ (assumed to be
distributed according to a uniform or a Gaussian distribution), (ii)
the transverse velocity has a direction angle $\Omega$ (assumed to be
uniformly distributed), and (iii) the system is found at its current
position in the Galaxy (determined by the time the system spends near
its current position divided by its kinematic age). 

Here, we determine the probability of finding the system at its
current position in the Galaxy in three dimensions (i.e., besides the
vertical distance to the Galactic plane, we also consider the radial
and azimuthal position in the plane). For this purpose, we determine
the time the system spends in a sphere with radius $R$ centered on its
current location for all $V_r$- and $\Omega$-values considered for the
derivation of the possible birth sites. The probability to find the
system near its current position is then determined as the time it
spends in this sphere divided by its age. For the latter, we adopt the
kinematic ages obtained by tracing the motion of the system backwards
in time, so that the probability of finding the system near its current
position depends on the radial velocity $V_r$, the proper motion
direction $\Omega$, and the Galactic plane crossing considered along
the orbit associated with $V_r$ and $\Omega$.  We performed some test
calculations to successfully verify that the results based on the thus
determined probabilities are insensitive to the adopted value of $R$,
as long as it is sufficiently small for the sphere to represent a
local neighborhood near the system's current position. For the results
presented in this paper we, somewhat arbitrarily, choose $R=50$\,pc.

 Ultimately, the probability that, for a given pair of $V_r$ and $\Omega$, the post-SN peculiar velocity of the binary is equal to $V_{\rm CM}$ is given by
\begin{equation}
P \left( V_{\rm CM}| V_r, \Omega \right) \propto 
  \sum_{i=1}^{N(V_r,\Omega)} {{ T \left( V_r, \Omega \right) } 
  \over {\tau_{{\rm kin},i} \left( V_r, \Omega \right)}}\, 
  \lambda_i \left( V_r, \Omega \right),  \label{pvcm}
\end{equation}
where $N(V_r,\Omega)$ is the number of Galactic plane crossings associated with $V_r$ and $\Omega$, $T(V_r, \Omega)$ is the time the system spends near its current position for the orbit associated with $V_r$ and $\Omega$, and $\tau_{{\rm kin},i} \left( V_r, \Omega \right)$ is the kinematic age corresponding to the $i$-th plane crossing associated with $V_r$ and $\Omega$. The factor $\lambda_i \left( V_r, \Omega \right)$ is equal to 1 when the $i$-th plane crossing along the orbit associated with $V_r$ and $\Omega$ yields a post-SN peculiar velocity equal to $V_{\rm CM}$, and equal to 0 otherwise. The total probability that the post-SN peculiar velocity of the binary is equal to $V_{\rm CM}$ is then obtained by integrating Eq.~(\ref{pvcm}) over all possible values of $V_r$ and $\Omega$:
\begin{equation}
P \left( V_{\rm CM} \right) \propto \int_{\Omega} \int_{V_r} 
  P \left( V_{\rm CM}| V_r, \Omega \right) P \left( V_r \right)
  P \left( \Omega \right) dV_r\, d\Omega.  \label{pvcm2}
\end{equation}
Here $P(V_r)$ is the probability distribution of the unknown radial velocity $V_r$, and $P(\Omega)$ the probability distribution of the unknown proper motion direction $\Omega$ in the plane perpendicular to the line-of-sight. 

For ages of 0-100\,Myr, post-SN peculiar velocities up to
100\,km\,s$^{-1}$ are likely for all four radial velocity
distributions. The highest post-SN peculiar velocities are found for
the Gaussian radial velocity distributions with velocity dispersions
of 130\,km\,s$^{-1}$ and 200\,km\,s$^{-1}$ ($V_{\rm CM}$ values remain
likely up to 200--300\,km\,s$^{-1}$) and the uniform radial velocity
distribution ($V_{\rm CM}$ values follow an almost flat distribution
from 300\,km\,s$^{-1}$ to 1500\,km\,s$^{-1}$). For ages of 49-51\,Myr,
the post-SN peculiar velocity distributions all have a peak at
50\,km\,s$^{-1}$. In the case of the uniform radial velocity
distribution and the Gaussian distributions with velocity dispersions
of 130\,km\,s$^{-1}$ or 200\,km\,s$^{-1}$, additional peaks of almost
equal height occur at even higher post-SN peculiar velocities. Similar
conclusions apply to the post-SN peculiar velocity distributions
obtained for a present-day transverse velocity component of
10\,km\,s$^{-1}$.  Despite the small lower limit on the present-day
{\em transverse} systemic velocity, a wide range of non-negligible post-SN
peculiar velocities is thus possible. We note however that the
distributions shown here incorporate all possible post-SN peculiar velocities
obtained from tracing the motion of the system backwards in time and
that some of these may require SN kicks and mass loss that are
incompatible with the observational constraints on the orbital
elements and component masses of the double pulsar.

\subsubsection{Progenitor constraints}
\label{progenitor}

After the formation of pulsar~B, the evolution of the system is
expected to be driven exclusively by the emission of gravitational
waves. We
use the differential equations derived by \citet{1992MNRAS.254..146J}
which are valid up to 3.5 post-Newtonian order of approximation, and 
 find the resulting post-SN orbital separation $A$
to be always between $1.26\, R_\odot$ and $1.54\, R_\odot$, and the
post-SN orbital eccentricity $e$ between $0.088$ and $0.12$. Since the kinematic ages are functions of the radial
velocity $V_r$, the proper motion direction $\Omega$, and the Galactic
plane crossing considered along the past orbit associated with $V_r$
and $\Omega$, the particular values of $A$ and $e$ associated with a
given $\tau_{\rm kin}$ are also functions of these variables.

The pre- and post-SN binary parameters and the kick velocity imparted
to pulsar~B at birth are related by the conservation laws of orbital
energy and angular momentum. For a circular pre-SN orbit, the
relations can be found in \citep{1983ApJ...267..322H,
1995MNRAS.274..461B, 1996ApJ...471..352K, 1997ApJ...489..244F, 
2000ApJ...530..890K}. 

The requirement that these equations permit real
solutions for the kick magnitude $V_k$, the kick orientation angles $\theta$, $\phi$, the NS progenitor mass $M_0$, and the pre-SN orbital semi-major axis $A_0$, imposes
constraints on all of these parameters. For a mathematical formulation of these constraints, we refer
to Eqs.~(21)--(27) in \citet{2005ApJ...625..324W}, and references
therein (a more compact description can also be found in \citet{2004ApJ...603L.101W} and \citet{2004ApJ...616..414W}. We here merely recall that the constraints express that: (i)
the binary components must remain bound after the SN explosion, (ii)
the pre- and post-SN orbits must pass through the instantaneous
position of the component stars at the time of the SN explosion, and
(iii) there is a lower and upper limit on the degree of orbital
contraction or expansion associated with a given amount of mass loss
and a given SN kick.

Besides the changes in the orbital parameters and the mass of the
exploding star, the SN explosion also imparts a kick velocity to the
binary's center of mass and tilts the post-SN orbital plane with
respect to the pre-SN one. The center-of-mass velocity and tilt
  angle add additional constraints on the solutions of the
  conservation laws of orbital energy and angular momentum.

The constraints on the progenitor of PSR\,J0737-3039 resulting from
the dynamics of asymmetric SN explosions arise solely from tracing the
evolution of the current system properties backwards in time. The
pre-SN orbital separations and pulsar~B progenitor masses found this
way are, however, not necessarily accessible through the currently
known DNS formation channels. Further constraints on the progenitor of
pulsar~B can therefore be obtained from stellar and binary evolution
calculations.

Firstly, a lower limit on the mass of pulsar~B's pre-SN progenitor is given by
the requirement that the star must be massive enough to evolve into a
NS rather than a white dwarf. According to our current understanding
of helium star evolution, the minimum helium star mass required for NS
formation is 2.1--2.8\,$M_\odot$ \citep{1986A&A...167...61H,
2003astro.ph..3456T}. The actual helium star minimum mass is, however,
still considerably uncertain due to the poorly understood evolution of
massive stars and possible interactions with close binary companions.

We impose a lower limit of $2.1\,M_\odot$ on
the mass of pulsar~B's pre-SN helium star progenitor. However, in the
light of recent suggestions that the progenitor of pulsar~B may have
been significantly less massive than the conventional lower limit of
$2.1\,M_\odot$, we also explore the possibility of progenitor masses
as low as pulsar B's present-day mass of $1.25\,M_\odot$
\citep{2005PhRvL..94e1102P, 2005MNRAS.361.1243P}. Unless our current
understanding of helium star evolution is seriously flawed, this
scenario implies that the progenitor of pulsar~B must have lost a
significant amount of mass (at least $\sim 0.7\,M_\odot$) after it had
already established a sufficiently massive core to guarantee the
occurrence of a future SN explosion.  We note that this possibility is included in the binary population synthesis calculations used to derive the adopted theoretical DNS radial-velocity distributions.

Secondly in \citet{2004ApJ...603L.101W}, we showed that the pre-SN binary was so tight ($1.2\,R_\odot \lesssim A_0 \lesssim 1.7\,R_\odot$) that the helium
star progenitor of pulsar~B must have been overflowing its Roche lobe
at the time of its SN explosion \citep[see also][]{2004MNRAS.349..169D}. An upper limit on the progenitor mass is
therefore given by the requirement that this mass-transfer phase be
dynamically stable (otherwise the components would have merged and no
DNS would have formed). Based on the evolutionary tracks for NS +
helium star binaries calculated by \citet{2003ApJ...592..475I}, we
adopt an upper limit of 3.5 on the mass ratio of the pre-SN
binary to separate dynamically stable from dynamically unstable
Roche-lobe overflow \citep[see also][]{2002MNRAS.331.1027D,
2003MNRAS.344..629D}.

\subsubsection{The most likely kick velocity and progenitor properties}
\label{mostlikely}

The most likely kick velocity and progenitor parameters obtained
from the procedure outlined in the previous sections are summarized in
Table~IV of \citep{2006PhRvD..74d3003W}. The most likely pulsar~B kick velocity is smaller
than 50\,km\,s$^{-1}$ only when the transverse velocity component has
a magnitude of 10\,km\,s$^{-1}$, the minimum helium star mass required
for NS formation is allowed to be as low as $1.25\,M_\odot$, and the
age of the system is assumed to be between 30 and 70\,Myr. {\em All
other models} yield most likely kick velocities of
50--180\,km\,s$^{-1}$. When a minimum pre-SN helium star mass of
$2.1\,M_\odot$ is imposed, the kicks are always strongly favored to be
directed opposite to the helium star's pre-SN orbital motion (most
likely $\cos \theta \simeq -0.90 \pm 0.05$). When the constraint on
the minimum pre-SN helium star mass is relaxed, the most likely kick
direction can shift significantly and can even become perpendicular to
the helium star's pre-SN orbital velocity. Allowing pre-SN helium star
masses down to $1.25\,M_\odot$ furthermore always leads to most likely
progenitor masses of $1.3-2.0\,M_\odot$, {\em except} when
$V_t=10\,{\rm km\,s^{-1}}$, the age of the system is between 0 and
100\,Myr, and the radial velocities are distributed uniformly or
according to a Gaussian with a velocity dispersion of
200\,km\,s$^{-1}$. In the latter cases, the most likely
  progenitor masses are 2.5--2.7\,$M_\odot$.

In summary, while small kick velocities of just a few tens of
km\,s$^{-1}$ could be favored for some models, the majority of the
models yields most likely values of 50--180\,km\,s$^{-1}$. Progenitor
masses below $2.1\,M_\odot$ are furthermore not required to explain
the system properties, although they are generally favored when helium
stars below $2.1\,M_\odot$ are still assumed to be viable NS
progenitors.  We note though that the results presented are based on the
assumption that all kick velocity magnitudes $V_k$ are equally
probable. This assumption is inconsistent for models using the
Gaussian radial velocity distributions based on population synthesis
calculations of coalescing DNSs. In particular, these calculations all
adopt a Maxwellian rather than a uniform kick velocity
distribution. In all cases, weighing the kick velocities according to
the Maxwellian underlying the derivation of the radial velocity
distributions would, however, shift the most likely kick velocities to
higher $V_k$ values. This reinforces
our conclusion that the presently known observational constraints not
necessarily disfavor kick velocity magnitudes of 100\,km\,s$^{-1}$ or
more.

\subsubsection{Confidence limits on the kick velocity and progenitor
  parameters}

\begin{figure*}
\resizebox{14.5cm}{!}{\includegraphics{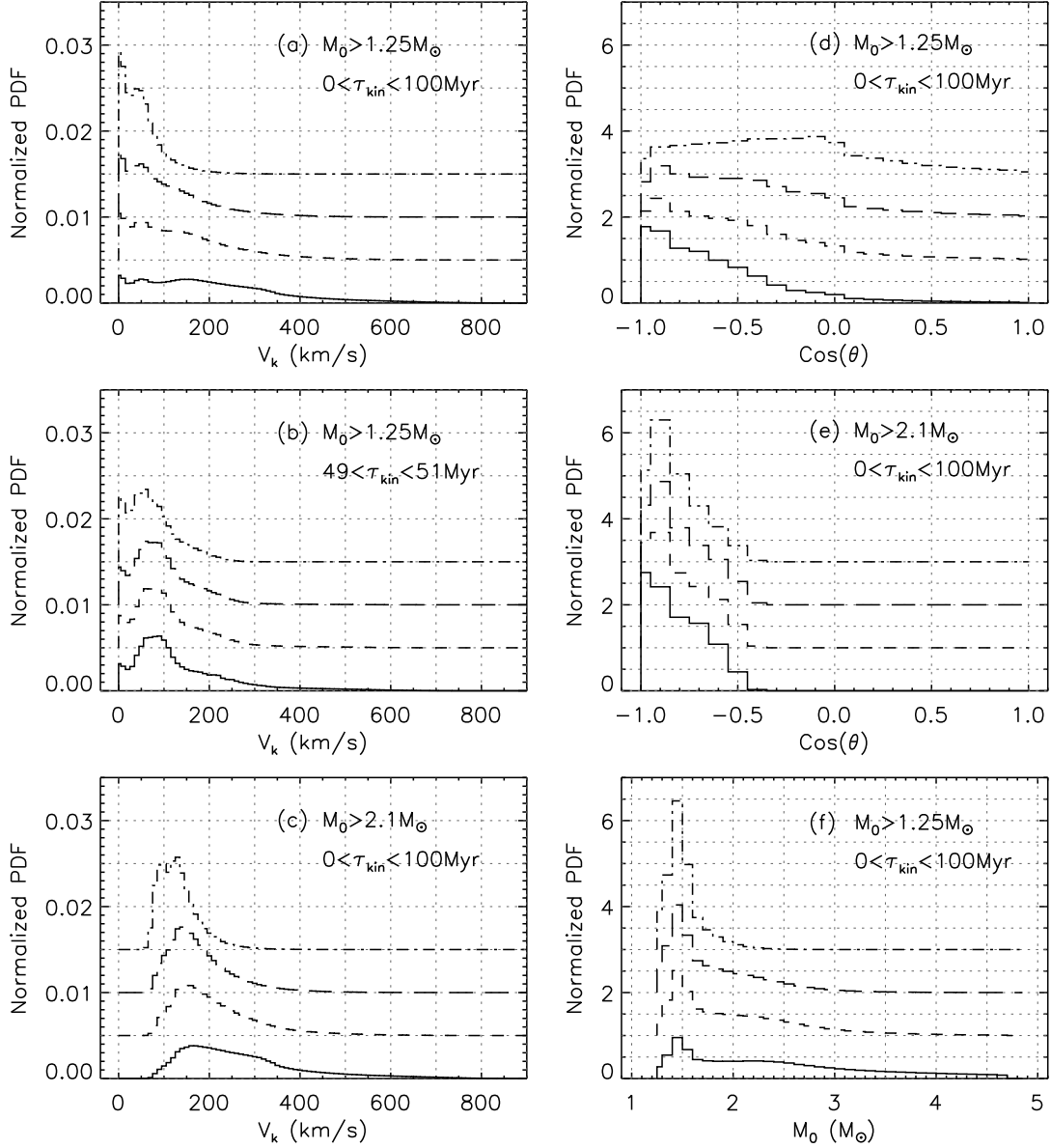}}
\caption{One-dimensional PDFs illustrating some of the dependencies of
  the derived pulsar~B kick velocity and progenitor properties on the
  adopted model assumptions. All plots are for a present-day
  transverse velocity of 10\,km\,s$^{-1}$. Solid lines correspond to
  uniform radial velocity distributions and dashed lines to Gaussian
  distributions with velocity dispersions of 60\,km\,s$^{-1}$
  (long-dashed lines), 130\,km\,s$^{-1}$ (short-dashed line), and
  200\,km\,s$^{-1}$ (dash-dotted line). For clarity, the PDFs are
  offset from each other by an arbitrary amount. Panels~(a)--(c) show
  the distributions of kick velocity magnitudes $V_k$, panels~(d)--(e)
  the distributions of kick direction cosines $\cos \theta$, and
  panel~(f) the distributions of pre-SN helium star masses $M_0$.} 
\label{1DPDF}
\end{figure*}

For illustration, some representative 1-D PDFs used for the
calculation of the confidence limits in the case of a present-day
transverse velocity component of 10\,km\,s$^{-1}$ are shown in
Fig.~\ref{1DPDF}. Panels~(a)--(c) show the kick velocity distributions
resulting from different present-day radial velocity distributions for
ages ranges of 0-100\,Myr and 49-51\,Myr, and minimum pre-SN helium
star masses of $1.25\,M_\odot$ and $2.1\,M_\odot$. For a given age
range and minimum pre-SN helium star mass, the PDFs show a peak which
is most pronounced when a Gaussian radial velocity distribution with a
velocity dispersion of 60\,km\,s$^{-1}$ is considered, and which
widens with increasing radial velocity dispersions. When the age is
assumed to be 0-100\,Myr and the minimum pre-SN helium star mass
$1.25\,M_\odot$, there is a clear tendency for the 1-D PDFs to favor
kick velocities of 50\,km\,s$^{-1}$ or less (although this becomes
significantly less pronounced with increasing radial velocity
dispersions). This trend shifts towards favoring kick velocities of
50--100\,km\,s$^{-1}$ when the age range is narrowed to
49-51\,Myr. Moreover, when the age range is kept fixed at 0-100\,Myr,
but the minimum pre-SN helium star mass is increased to
$2.1\,M_\odot$, the favored range of kick velocities shifts to
100--150\,km\,s$^{-1}$. In the latter case, the kick velocity is
furthermore always required to be larger than $\sim 60\,{\rm
km\,s^{-1}}$ \citep[see also][]{2004ApJ...603L.101W, 2004ApJ...616..414W}.

Panel~(f), finally, shows the distribution of possible pre-SN
progenitor masses for different present-day radial velocity
distributions, DNS ages of 0-100\,Myr, and a minimum pre-SN
helium star mass of $1.25\,M_\odot$. The distributions all favor
progenitor masses of $1.4$--$1.5\,M_\odot$, with the preference for
this mass range being strongest for present-day radial velocity
distributions with small radial velocity dispersions. Distribution
functions for a minimum pre-SN helium star mass of $2.1\,M_\odot$ look
practically the same as the ones displayed panel~(f) if they were cut
off at $2.1\,M_\odot$.

\subsubsection{Summary and concluding remarks}

One of the results of the above analysis is that marginalizing the full five-dimensional progenitor PDF to 1-D or 2-D distributions for the pulsar~B kick velocity and progenitor mass has a major effect on the determination of their most likely values. When the full multi-dimensional PDF is examined, it is clear that although some sets of prior assumptions indeed favor low kick velocities and low progenitor masses the majority of the models favor kick velocities of 50--180\,km\,s$^{-1}$ and progenitor masses of 1.45--2.75\,$M_\odot$. 

In particular, if the transverse systemic velocity is assumed to be 10\,km/s and helium stars less massive than 2.1\,$M_\odot$ are assumed to be viable NS progenitors, the most likely progenitor mass can vary from  1.35\,$M_\odot$ to 2.65\,$M_\odot$, depending on the assumed systemic age and radial velocity distribution. Hence, whether progenitor masses greater than 2.1\,$M_\odot$ are statistically likely or unlikely depends strongly on the adopted prior assumptions \citep[cf.][]{2005PhRvL..94e1102P, 2005astro.ph.10584P}. Most likely progenitors with $M_0 < 2.1\,M_\odot$ can furthermore also be associated with kick velocities of up to 100\,km\,s$^{-1}$. 
\citet{2006MNRAS.tmpL.105S} recently arrived at similar conclusions based on
  updated proper motion measurements by \citet{2006Sci...314...97K}. The
  updated measurements yield a proper motion of 10\,km/s in a
  direction nearly parallel to the plane of the Galaxy.

We also find that the proximity of PSR\,J0737-3039 to the
Galactic plane and the small proper motion do not pose stringent
constraints on the kick velocity and progenitor mass of
pulsar~B. Instead, the constraints are predominantly determined by the
orbital dynamics of asymmetric SN explosions.  This is in contrast to the conclusions of \citet{2005PhRvL..94e1102P} and \citet{2005astro.ph.10584P} who emphasize that the proximity of the double pulsar to the Galactic plane implies that pulsar~B most likely received only a small kick at birth and that its progenitor most likely had mass of  $\sim 1.45\,M_\odot$.

Hence, based on the currently available observational constraints, a
wide range of progenitor and kick velocity properties are favored for
PSR\,J0737-3039B. In particular, the constraints are compatible with a
conventional hydrodynamical or neutrino-driven SN explosion from a
helium star more massive than $2.1\,M_\odot$
\citep{1986A&A...167...61H, 2003astro.ph..3456T}, as well as an
electron-capture SN from a helium star less massive than
$2.5\,M_\odot$ \citep{1984ApJ...277..791N,
1987ApJ...322..206N}. \citet{2005MNRAS.361.1243P} have speculated that
the electron-capture SN mechanism may be typical for close binaries
and that it may be accompanied by much smaller kicks than
hydrodynamical or neutrino-driven SN explosions. Consequently, if
pulsar~B is formed through an electron capture SN and if electron
capture SNe are accompanied by small kicks, the mass of pulsar~B's
pre-SN helium star progenitor must be smaller than $2.1\,M_\odot$
(otherwise the kick is always larger than $\sim 60\,{\rm
km\,s^{-1}}$). Since it is unlikely that future observations will lead to new constraints on the smallest possible pulsar~B progenitor mass,
further insight to the formation mechanism of pulsar~B should be
sought in core-collapse simulations of low-mass ($\lesssim
2.1\,M_\odot$) helium stars and population synthesis studies of
PSR\,J0737-3039-type binaries and their progenitors.

\subsection{PSR~B1534+12}

The relativistic binary radio pulsar
PSR\,B1534+12 too has an accurately measured proper motion with a known
direction in right ascension and declination,
so that its kinematic history and progenitor constraints {\em depend
only on the unknown radial velocity $V_r$}. In order to derive these
constraints, we adopt the spin-down age $\tau_b=210$\,Myr as an upper
limit to the age of the system. The other physical parameters of
PSR\,B1534+12 relevant to the derivation are summarised in
Table~1 of \citet{2004ApJ...616..414W}.

Following the same arguments as for PSR\,J0737-3039, we assume that the progenitor of PSR\,B1534+12 was close to the
Galactic plane at the time of the second SN explosion and that its
pre-SN systemic velocity was almost entirely due to Galactic
rotation. The possible birth sites of the DNS are then obtained by
tracing the Galactic motion of the system backwards in time as a
function of the unknown radial velocity $V_r$. We find that, within
the imposed age limit of 210\,Myr, no Galactic plane crossings occur
when $V_r \gtrsim 250$\,km\,s$^{-1}$, whereas up to four crossings may
occur when $V_r \lesssim 250$\,km\,s$^{-1}$. Since four disk crossings only
occur for relatively few and rather fine-tuned Galactic trajectories,
we leave these cases aside and focus on the possible birth sites
associated with the first (case~A), second (case~B), and third
(case~C) Galactic plane crossings.

\begin{figure}
\begin{center}
\resizebox{\hsize}{!}{\includegraphics{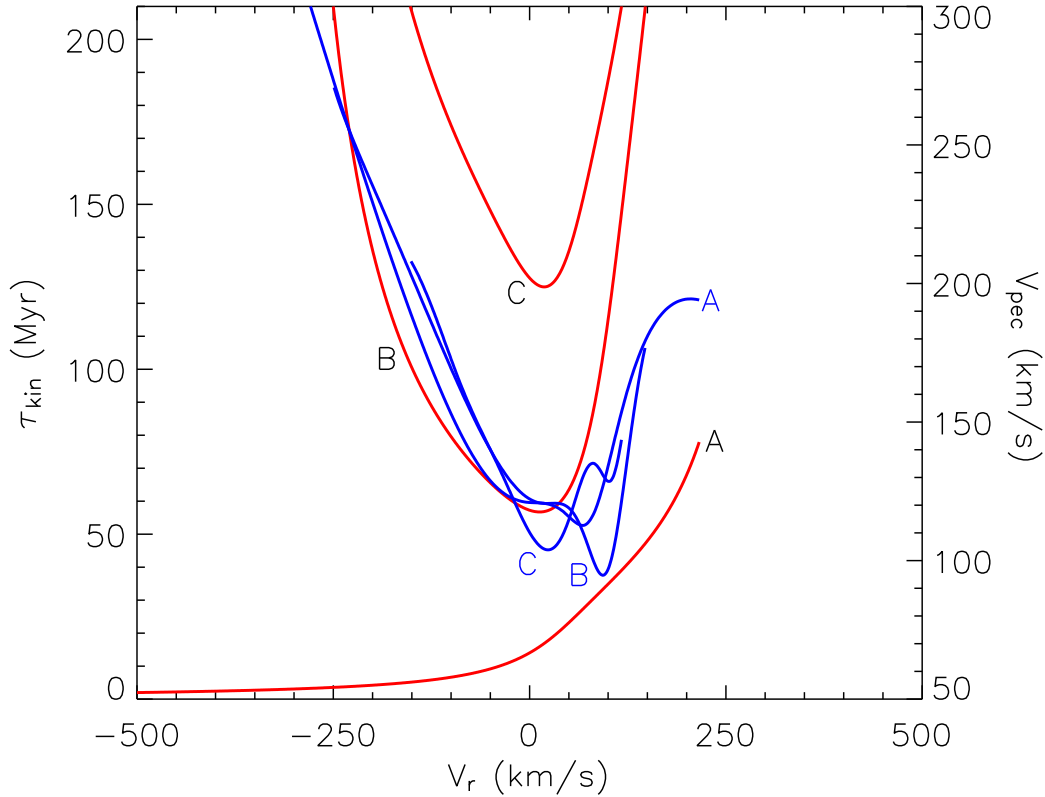}}
\end{center}
\caption{Variations of the kinematic age (left-hand axis)
  and post-SN peculiar velocity (right-hand axis) of
  PSR\,B1534+12 as a function of the unknown radial velocity $V_r$,
  for cases~A, B, and~C. The kinematic ages are represented by red lines and the peculiar
  velocities by blue lines.}  
\label{tauk4}
\end{figure}

The kinematic ages and post-SN peculiar velocities associated with the
Galactic plane crossings are shown in Fig.~\ref{tauk4} as functions
of the radial velocity $V_r$. Case~A gives rise to a wide range of
ages between $\simeq 1$\,Myr and 210\,Myr, and peculiar velocities
between 110\,km\,s$^{-1}$ and 1500\,km\,s$^{-1}$. Cases~B and C, on
the other hand, yield kinematic ages of {\em at least} $\simeq
55$\,Myr and $\simeq 125$\,Myr; and post-SN peculiar velocities of
90--270\,km\,s$^{-1}$ and 100--220\,km\,s$^{-1}$, respectively.

\begin{figure*}
\resizebox{\hsize}{!}{\includegraphics{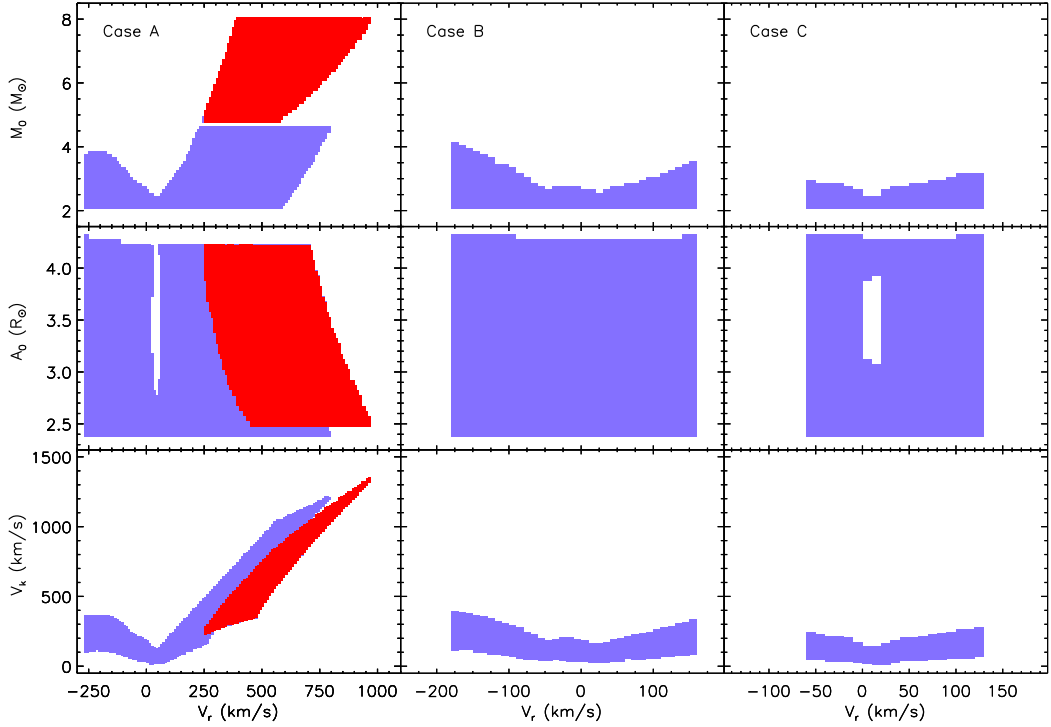}}
\caption{Limits on the pre-SN progenitor of PSR\,B1534+12 and on the
  kick velocity imparted to the last-born NS. The left-hand panels
  show the constraints for case~A, the middle panels for case~B, and
  the right-hand panels for case~C. Red regions correspond to solutions for which the pre-SN binary
  is detached, while blue 
  regions indicate the additional solutions associated with
  mass-transferring systems.}
\label{pro10}
\end{figure*}

The post-SN orbital separation $A$ and orbital eccentricity $e$ at the
times of Galactic disk crossings are obtained by numerical integration
backwards in time of the equations governing the evolution of the
orbit under the influence of gravitational radiation. The resulting
post-SN orbital parameters range from $A=3.28\,R_\odot$ and $e=0.274$
when $\tau_{kin}=0$\,Myr to $A=3.36\,R_\odot$ and $e=0.282$ when
$\tau_{kin}=210$\,Myr. These post-SN orbital parameters together with
the post-SN peculiar velocities impose constraints on the pre-SN
progenitor of PSR\,B1534+12. These are derived in a similar way as for
the progenitor of PSR\,J0737-3037 (except that the problem depends on
only one free parameter, $V_r$). The results of our analysis are
summarized in Fig.~\ref{pro10}.

Based on the observational constraints available in 2004, PSR\,B1534+12 may have
been detached as well as semi-detached at the time the second NS was
born. In order to illustrate this, the solutions for which no mass
transfer takes place at the time of the second SN explosion are
indicated by the red regions in
Fig.~\ref{pro10}, while the blue regions indicate the {\it additional} solutions that become
accessible when the possibility of mass transfer is taken into
account. For the latter solutions we adopt the same assumption as
before that the system may survive the mass transfer phase and form a
DNS only if the mass ratio is smaller than 3.5 \citep[e.g.,][]{2003ApJ...592..475I}.

For case~A disk crossings, detached pre-SN binary configurations
exist only for $V_r \lesssim-250$\,km\,s$^{-1}$. For larger radial velocities,
the system is always undergoing mass transfer from the progenitor of
the second-born NS to the first-born NS. In this analysis, the pre-SN mass of the helium
star forming the second NS is constrained within $2.1\,M_\odot - 8\,M_\odot$. The lower limit again corresponds to the lowest mass for which a helium star is expected to form a NS instead of a white dwarf, while the upper limit corresponds to the highest mass for which a helium star is expected to form a NS rather than a black hole (see, e.g., Fig.~1 in \citealt{2002ApJ...572..407B} and Table~16.4 in \citealt{2003astro.ph..3456T}). The divide
between the red and blue regions at $4.7\,M_\odot$ corresponds 
to the adopted critical mass ratio of $3.5$ for dynamically
stable mass transfer. The allowed mass range of pre-SN helium star
masses is most constrained for $|V_r| \lesssim 200$\,km\,s$^{-1}$ when
$2.1\,M_\odot \lesssim M_0 \lesssim 4\,M_\odot$. Lower and upper limits on the
pre-SN orbital separation are given by $2.4\,R_\odot$ and
$4.3\,R_\odot$. 

\begin{figure*}
\begin{center}
\resizebox{\hsize}{!}{\includegraphics{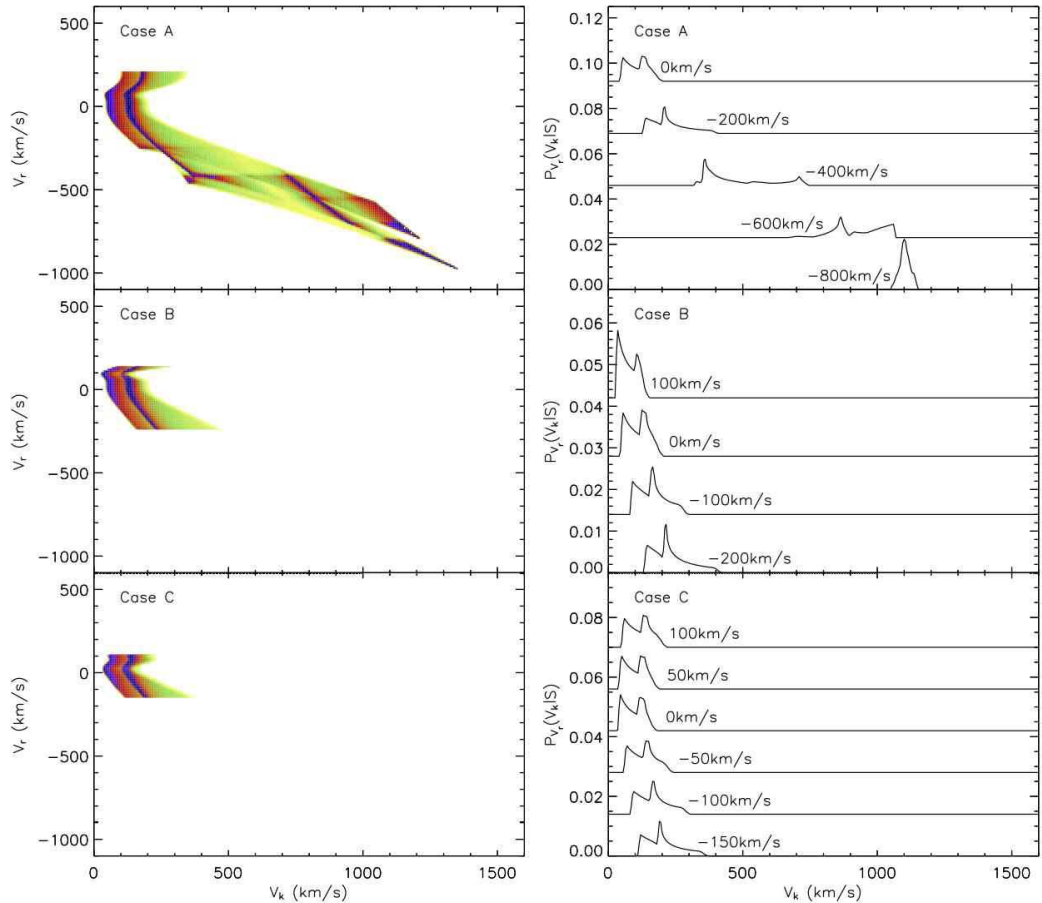}}
\end{center}
\caption{Probability distribution functions of the magnitude of the
kick velocity imparted to PSR\,B1534+12's companion at the time of its
formation, for cases~A, B, and~C. The left panels show the entire set of PDFs associated with all admissible Vr-values by means of a linear color scale that varies over yellow, green, orange, red, and blue with increasing PDF values, while the right panels show the PDFs associated with some specific Vr-values. For clarity, the curves in the right panels are offset from each other by an arbitrary amount. } 
\label{vk4}
\end{figure*}

The behavior of the kick-velocity magnitude is similar as for
PSR\,J0737-3039: the kick velocity increases with increasing absolute
values of $V_r$ and, for a given radial velocity, it is generally
constrained to an interval that is less than $\simeq
600$\,km\,s$^{-1}$ wide. Solutions for which the pre-SN helium star is
overflowing its Roche lobe furthermore allow for somewhat higher kick
velocities than solutions for which the helium star fits within its
critical Roche lobe. These higher kick velocities are associated with
the smaller orbital separations that become accessible for
mass-transferring progenitors. When all possible pre-SN binary
configurations are considered, the kick velocity magnitudes range from
45\,km\,s$^{-1}$ to 1350\,km\,s$^{-1}$. When configurations for which
the helium star is overflowing its Roche lobe are excluded, the range
narrows to $230\,{\rm km\,s^{-1}} \lesssim V_k \lesssim 1350\,{\rm km\,s^{-1}}$,
so that the minimum required kick velocity is higher than when the
possibility of Roche-lobe overflow is taken into account. The larger
minimum kick velocity is required to compensate the larger effect of
the mass lost from the system during the SN explosion: since the
minimum $M_0$ for these systems is $4.7\,M_\odot$, at least $57\%$ of
the total pre-SN mass is lost from the system.

\begin{figure*}
\begin{center}
\resizebox{\hsize}{!}{\includegraphics{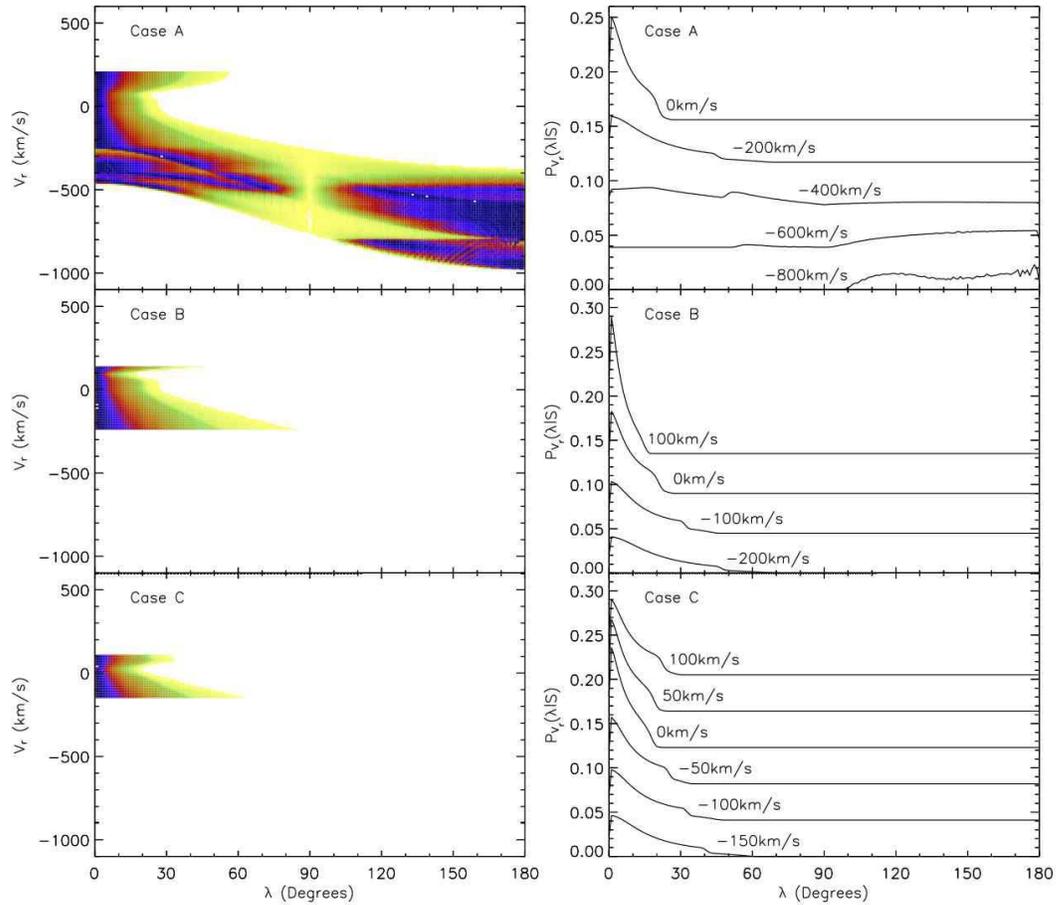}}
\end{center}
\caption{Probability distribution functions for the misalignment angle
$\lambda$ between PSR\,B1534+12's spin axis and the post-SN orbital
angular momentum axis, for cases~A, B, and~C (cf. Fig.~\ref{vk4} for
more details).}
\label{tilt3}
\end{figure*}

For case~B disk crossings, almost all allowed pre-SN binary
configurations imply that the helium star progenitor of the
second-born NS is overflowing its Roche lobe at the time of its SN
explosion. The mass of the helium star is constrained to $2.1\,M_\odot
\lesssim M_0 \lesssim 4.9\,M_\odot$, the pre-SN orbital separation to
$2.4\,R_\odot \lesssim A_0 \lesssim 4.3\,R_\odot$, and the kick-velocity
magnitude to $30\,{\rm km\,s^{-1}} \lesssim V_k \lesssim 475\,{\rm
km\,s^{-1}}$. If
the helium star fits within its critical Roche lobe at the time of its
SN explosion, the constraints become $M_0 \approx 4.8\,M_\odot$, $A_0
\approx 4.2\,R_\odot$, $V_k \approx 240\,{\rm km\,s^{-1}}$.

Case~C disk crossings, finally, yield no detached solutions for the
progenitor of PSR\,B1534+12. The constraints for this case are similar
to those found for case~B, except that the mass $M_0$ is always
smaller than $\simeq 4\,M_\odot$ and the kick velocity $V_k$ is always
smaller than $370\,{\rm km\,s^{-1}}$.

The constraints derived above may again be used to derive probability distribution functions for the magnitude $V_k$ of the kick velocity imparted to the second-born NS and for the misalignment angle $\lambda$ between the spin-axis of the first-born NS and the post-SN orbital angular momentum axis. 
We note that, unlike the PDFs derived for the double
pulsar, the PDFs presented in this section do not incorporate probability
distributions for $V_r$ obtained from theoretical population synthesis
calculations, nor do they incorporate the probability of finding the
system at its current place in the Galaxy. 
The resulting PDFs are presented in Figs.~\ref{vk4} and~\ref{tilt3}. For radial velocities of only a few 100\,km\,s$^{-1}$, the kick-velocity distributions show two closely spaced and fairly evenly matched peaks between $V_k \simeq 50$\,km\,s$^{-1}$ and $V_k \simeq 250$\,km\,s$^{-1}$. For higher radial velocities, relevant only to case~A, the peak(s) shift to larger kick velocities up to a maximum of $\simeq 1350$\,km\,s$^{-1}$. The tilt-angle distributions, on the other hand, favor misalignment angles below $30^\circ$ when $|V_r| \lesssim 200$\,km\,s$^{-1}$ and above below $100^\circ$ when $|V_r| \gtrsim 600$\,km\,s$^{-1}$. For case~A disk crossings, tilt angles close to $\lambda \approx 90^\circ$ are furthermore strongly disfavored regardless of the value of the radial velocity.  For case~B  and C disk crossings, tilt angles with non-vanishing probabilities are always smaller than $40^\circ$--$50^\circ$. For kick velocities of less
than 200\,km\,s$^{-1}$, the predictions in all three cases are in
good agreement with the measurement of $\lambda = 25.0 \pm 3.8^\circ$
by \citet{2004PhRvL..93n1101S}.

We conclude by noting that updated progenitor constraints by \citet{2005ApJ...619.1036T} incorporating the tilt angle measured by \citet{2004PhRvL..93n1101S} exclude the possibility that the progenitor was detached
right before the SN explosion forming the second NS.  

\subsection{PSR~B1913+16: The Hulse-Taylor Binary Pulsar}

Besides the knowledge of the proper motion direction, PSR\,B1913+16
has the advantage that the misalignment angle between the pulsar's
spin axis and the pre-SN orbital angular momentum axis has been
determined to be around $\simeq 20^\circ$, corresponding to prograde
rotation, or $\simeq 160^\circ$, corresponding to retrograde rotation
\citep{1998ApJ...509..856K, 2002ApJ...576..942W}. \citet{2000ApJ...528..401W} used this
information to derive constraints on the mass of the second-born NS's
direct progenitor, on the pre-SN orbital separation, and on the
magnitude and direction of the kick velocity imparted to the
second-born NS at birth. In agreement with the then available
numerical simulations of rapidly accreting neutron stars, the authors
assumed that mass transfer from a helium star companion would cause
the NS to collapse into a black hole and therefore excluded Roche-lobe
overflowing helium stars as viable progenitors of the second-born
NS. However, more recent calculations by \citet{2002MNRAS.331.1027D}, \citet{2003ApJ...592..475I}, and \citet{2003MNRAS.344..629D} show that NS binaries may
survive a helium-star mass-transfer phase, provided that the ratio of
the helium star's mass to the neutron star's mass is not too extreme
($\le 3.5$). We therefore revise the
constraints derived by \citet{2000ApJ...528..401W} in the light of this new
information.  The constraints on the tilt angle, for which we
consider the values $\lambda=18^\circ \pm 6^\circ$ and
$\lambda=162^\circ \pm 6^\circ$ are also imposed.

We look for possible birth sites of PSR\,B1913+16 by tracing the
system's motion in the Galaxy backwards in time up to a maximum age of
80\,Myr\footnote{As for PSR\,J0737-3039 and PSR\,B1534+12, we use the
spin-down age $\tau_b=80$\,Myr instead of the characteristic age
$\tau_c=110$\,Myr as an upper limit for the age of the system. The
maximum amount of orbital evolution that may have taken place since
the formation of the DNS is therefore somewhat smaller than in \citet{2000ApJ...528..401W}.}, as a function of the unknown radial velocity $V_r$. In
agreement with \citet{2000ApJ...528..401W}, we find that the system may have
crossed the Galactic plane up to two times. The first Galactic plane
crossing (case~A) occurs at very young kinematic ages of $\simeq
2$--4\,Myr and gives rise to peculiar velocities in excess of $\simeq
300$\,km\,s$^{-1}$. The corresponding post-SN orbital parameters are
$A \approx 2.8\, R_\odot$ and $e \approx 0.618$. The second Galactic
plane crossing (case~B), on the other hand, takes place at least
$\simeq 55$\,Myr in the past and yields peculiar velocities of $\simeq
230$--440\,km\,s$^{-1}$. The associated post-SN orbital separations
and eccentricities range from $A=3.1\,R_\odot$ and $e=0.646$ to
$A=3.2\,R_\odot$ and $e=0.658$.

\begin{figure*}
\begin{center}
\resizebox{\hsize}{!}{\includegraphics{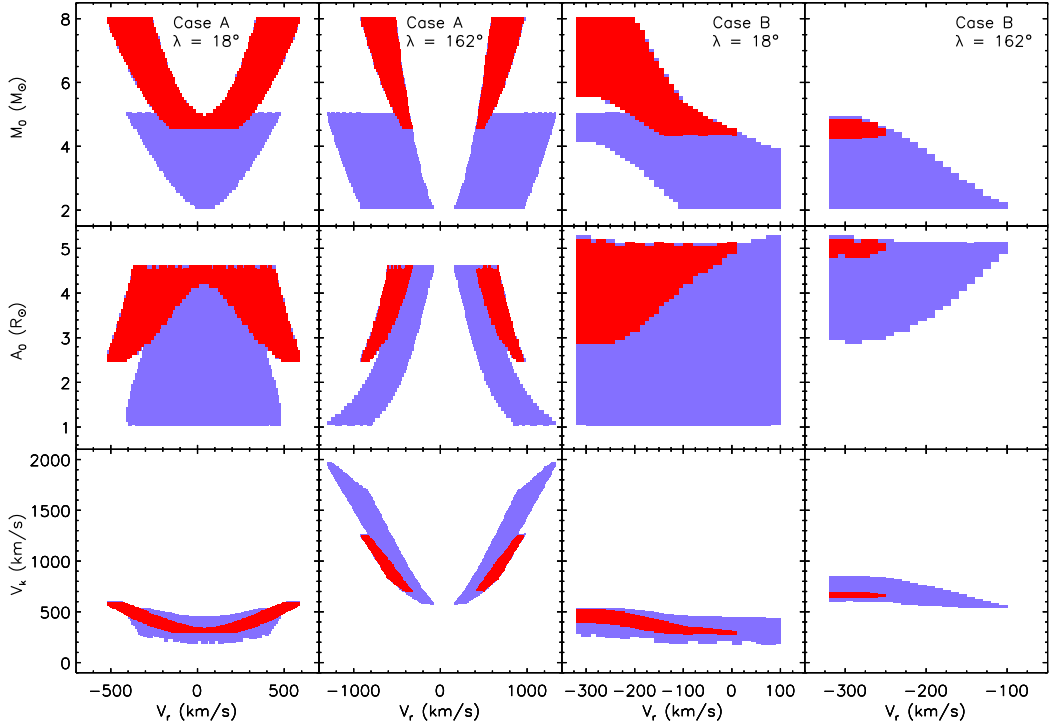}}
\end{center}
\caption{Limits on the pre-SN progenitor of PSR\,B1913+16 and on the
  kick velocity imparted to the last-born NS. Red regions correspond to solutions for which the  pre-SN binary is detached, while blue  regions indicate the additional solutions associated with
  mass-transferring systems.}
\label{pro9}
\end{figure*}

The constraints on the pre-SN parameter space accessible to the
progenitor of PSR\,B1913+16 are shown in Fig.~\ref{pro9} as functions
of the unknown radial velocity $V_r$. As before, the red regions correspond to the solutions for which
no mass transfer takes place at the time of the second SN explosion,
while the blue regions indicate
the additional solutions found when the possibility of mass transfer
is taken into account. The constraints for detached pre-SN binary
configurations are in good agreement with the constraints derived by
\citet{2000ApJ...528..401W}. It is clear however, that when the possibility of
mass transfer is taken into account, the available parameter space
becomes much less constrained.

For case~A disk crossings, the range of radial velocities for which
physically acceptable solutions exist is restricted to $|V_r| \lesssim
500$--600\,km\,s$^{-1}$ when $\lambda=18^\circ$, and to $100\,{\rm
km\,s^{-1}} \lesssim |V_r| \lesssim 1300\,{\rm km\,s^{-1}}$ when
$\lambda=162^\circ$. The mass of the second-born NS's direct
progenitor is constrained to the interval between $2.1\,M_\odot$ and
$8\,M_\odot$, and the pre-SN orbital separation to the interval
between $1.1\,R_\odot$ and $5.3\,R_\odot$ for both considered values
of the tilt angle $\lambda$. The magnitude of the kick velocity varies
from $\simeq 190$\,km\,s$^{-1}$ to $\simeq 600$\,km\,s$^{-1}$ when
$\lambda=18^\circ$ and from $\simeq 580$\,km\,s$^{-1}$ to $\simeq
2000$\,km\,s$^{-1}$ when $\lambda=162^\circ$. When the solutions are
restricted to detached pre-SN binary configurations the limits become
$300\,{\rm km\,s^{-1}} \lesssim V_k \lesssim 600\,{\rm km\,s^{-1}}$ and
$700\,{\rm km\,s^{-1}} \lesssim V_k \lesssim 1250\,{\rm km\,s^{-1}}$ for
$\lambda=18^\circ$ and $\lambda=162^\circ$, respectively.

For case~B disk crossings, the mass $M_0$ of the second-born NS's
direct progenitor and the pre-SN orbital separation $A_0$ are
constrained to the ranges of values given by $2.1\,M_\odot \lesssim M_0 \lesssim
8\,M_\odot$ and $1.1\,R_\odot \lesssim A_0 \lesssim 5.3\,R_\odot$ when
$\lambda=18^\circ$, and to $2.1\,M_\odot \lesssim M_0 \lesssim
5\,M_\odot$ and $2.9\,R_\odot \lesssim A_0 \lesssim 5.3\,R_\odot$ when
$\lambda=162^\circ$. The magnitude of the kick velocity varies between
$\simeq 190$\,km\,s$^{-1}$ and $\simeq 530$\,km\,s$^{-1}$ when
$\lambda=18^\circ$, and between $\simeq 550$\,km\,s$^{-1}$ and $\simeq
850$\,km\,s$^{-1}$ when $\lambda=162^\circ$. When the solutions are
restricted to those where no Roche-lobe overflow occurs at the time of
the helium star's SN explosion, the minimum kick velocity associated
with $\lambda=18^\circ$ increases slightly to approximately
280\,km\,s$^{-1}$, while the range of admissible kick velocities
associated with $\lambda=162^\circ$ becomes very tightly constrained
to $V_k \approx 640$--680\,km\,s$^{-1}$.

\begin{figure*}
\begin{center}
\resizebox{\hsize}{!}{\includegraphics{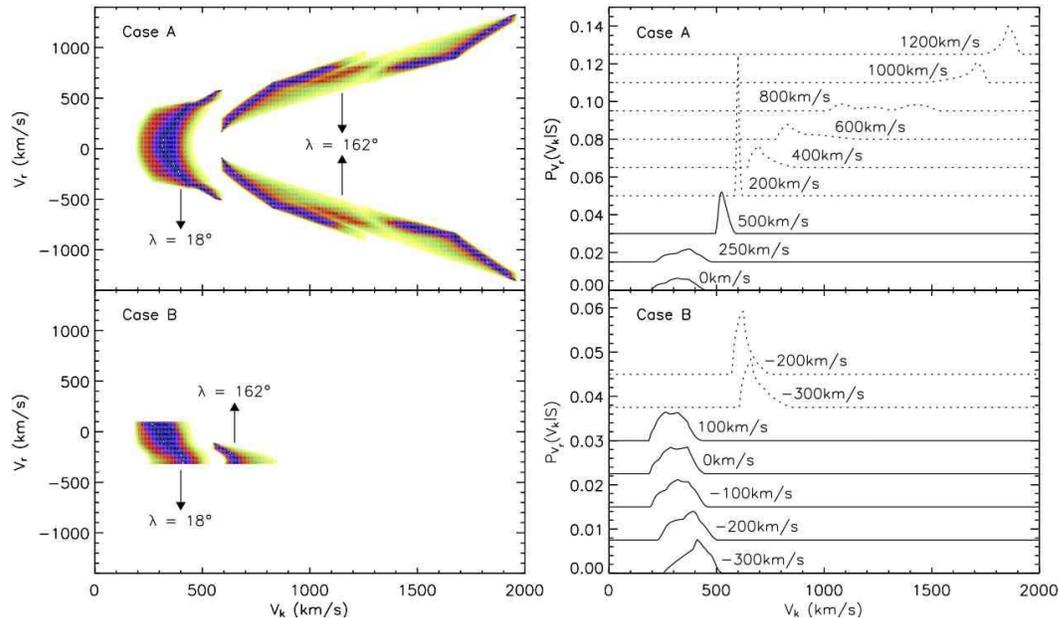}}
\end{center}
\caption{Probability distribution functions of the magnitude of the
kick velocity imparted to PSR\,B1913+16's companion at the time of its
formation, for both $\lambda=18^\circ$ and $\lambda=162^\circ$
(cf. Fig.~\ref{vk4} for more details). In the right-hand panels, the
solid lines correspond to $\lambda=18^\circ$ and the dashed lines to
$\lambda=162^\circ$.}
\label{vk2}
\end{figure*}

The probability distribution functions for the magnitude of the kick
velocity imparted to the second-born NS in PSR\,B1913+16 are shown in
Fig~\ref{vk2} for both $\lambda=18^\circ$ and $\lambda=162^\circ$. The
distributions generally show a single peak at kick velocities which
increase with increasing absolute values of $V_r$. For
$\lambda=18^\circ$, the most probable kick velocity ranges from 
$\simeq 300$\,km\,s$^{-1}$ to $\simeq 600$\,km\,s$^{-1}$, while for
$\lambda=162^\circ$ it ranges from $\simeq 600$\,km\,s$^{-1}$ to
almost $\simeq 2000$\,km\,s$^{-1}$.

\section{Formation Channels for Double Compact Objects}
\label{channels}

The inspiral and coalescence of double compact objects (DCO), such as NS-NS, BH-NS, and BH-BH binaries are some of the most promising
candidate events for GW detection by current ground-based interferometers, like LIGO. The formation of double compact object populations has been studied by a number of different groups using population synthesis techniques. In one of these studies \citep{2002ApJ...572..407B} we focused on an investigation of the main formation channels, their origin, and their relative contributions. The details of this study the synthesis code used ({\em StarTrack}) and the conclusions are presented in \citet{2002ApJ...572..407B}. Here we merely summarize the findings related to the various formation channels and their relative importance.  Results are described for our standard model from that study (model A described in \S\,2.1 and \S\,2.2 of \citealt{2002ApJ...572..407B}). 

We consider double compact objects with NS or BH (NS-NS, BH-NS or
BH-BH binaries) with merger times shorter than $10^{10}$\ yr.  In our standard evolutionary model, the population of
coalescing DCO is dominated by NS-NS systems (61\%), with a significant
contribution by BH-BH binaries (30\%), and a small contribution by BH-NS
objects (9\%). 

In Table~\ref{channels}, we present the most important formation
channels of coalescing DCO, for our standard model.  Formation channels
of NS-NS, BH-NS and BH-BH binaries are marked by NSNS, BHNS and BHBH,
respectively, they are listed in order of decreasing relative formation
frequency (second column) with respect to the whole DCO coalescing
population.  The details of each evolutionary sequence, i.e., MT
episodes and SN explosions are also given.  Results were obtained based
on the evolution of $3\times 10^7$ primordial binaries.

\subsection{Populations of Double Neutron Stars}

\citet{2002ApJ...572..407B} identified a
number of new NS-NS formation channels.  This is a result of
two improvements in the implementation of our population synthesis code,
since \citealt{2001ApJ...550L.183B}, hereafter BK01.  First, we have replaced the
approximate prescription suggested by \citet{1998ApJ...506..780B} for the
hyper-critical accretion during CE phases, with a newly derived
numerical solution (see Appendix of \citealt{2002ApJ...572..407B}). Second, we allow for hyper-critical
CE evolution of low-mass helium stars with compact objects. In BK01, we allowed binaries with low-mass helium
giants to evolve through DCE and standard SCE, but we had assumed that
CE events of helium giants with compact objects lead to mergers, and
possibly a gamma-ray burst (e.g., \citealt{1998ApJ...496..333F}). However, due to  the small mass of helium giant envelope at the onset of CE event ($\sim
1-1.5 M_\odot$), we find that these systems survive the CE events, and
form very tight NS-NS binaries.

Double neutron stars are formed in various ways through more than 14
different evolutionary channels identified in Table~\ref{channels}. The relative formation efficiencies shown for each channel are for the standard model A, described in \citet{2002ApJ...572..407B}. 
The entire population of coalescing NS-NS systems, may be
divided into three main subgroups.

{\em Group I.} This subpopulation consists of non-recycled NS-NS
systems, first identified by BK01. These are systems in which none of
the two NS ever had a chance of getting recycled through accretion.  
Our current results for the predicted formation rates and properties of
the non-recycled NS-NS systems, have not been affected by the two
improvements discussed above.  As shown 

\begin{tabular}[t]{ccc}

& &  Double Compact Object Formation Channels   \\

\hline

Formation & Relative   & \\
Channel & Efficiency$^{\rm \alpha}$ & Evolutionary History$^{\rm \beta}$  \\

\hline

NSNS:01& 20.3\ \%& 
NC:a$\rightarrow$b, SN:a, HCE:b$\rightarrow$a, HCE:b$\rightarrow$a, SN:b\\

NSNS:02& 10.8\ \%&
NC:a$\rightarrow$b, SCE:b$\rightarrow$a, NC:a$\rightarrow$b, SN:a, 
HCE:b$\rightarrow$a, SN:b\\

NSNS:03& 5.5\ \%& 
SCE:a$\rightarrow$b, SN:a, HCE:b$\rightarrow$a, HCE:b$\rightarrow$a, SN:b\\

NSNS:04& 4.0\ \%&
NC:a$\rightarrow$b, SCE:b$\rightarrow$a, SCE:b$\rightarrow$a, SN:b,
HCE:a$\rightarrow$b, SN:a\\

NSNS:05& 3.2\ \%&
DCE:a$\rightarrow$b, SCE:a$\rightarrow$b, SN:a, HCE:b$\rightarrow$a, SN:b\\

NSNS:06& 2.5\ \%&
SCE:a$\rightarrow$b, SCE:b$\rightarrow$a, NC:a$\rightarrow$b, SN:a,
HCE:b$\rightarrow$a, SN:b\\

NSNS:07& 2.2\ \%&
NC:a$\rightarrow$b, NC:a$\rightarrow$b, SN:a, HCE:b$\rightarrow$a,
HCE:b$\rightarrow$a, SN:b\\

NSNS:08& 2.0\ \%& 
NC:a$\rightarrow$b, DCE:b$\rightarrow$a, SN:a, HCE:b$\rightarrow$a, SN:b\\

NSNS:09& 2.0\ \%&
DCE:a$\rightarrow$b, DCE:a$\rightarrow$b, SN:a, SN:b\\

NSNS:10& 1.6\ \%&
NC:a$\rightarrow$b, SCE:b$\rightarrow$a, SN:b, HCE:a$\rightarrow$b, SN:a\\

NSNS:11& 1.5\ \%&
NC:a$\rightarrow$b, SCE:b$\rightarrow$a, DCE:b$\rightarrow$a, SN:a, SN:b\\

NSNS:12& 1.5\ \%&
NC:a$\rightarrow$b, SCE:b$\rightarrow$a, DCE:a$\rightarrow$b, SN:a, SN:b\\

NSNS:13& 1.0\ \%&
DCE:a$\rightarrow$b, SN:a, HCE:b$\rightarrow$a, SN:b\\

NSNS:14& 3.0\ \%&
all other \\

&&\\

BHNS:01& 4.5\ \%&
NC:a$\rightarrow$b, SN:a, HCE:b$\rightarrow$a, SN:b\\

BHNS:02& 1.6\ \%&
NC:a$\rightarrow$b, SCE:b$\rightarrow$a, SN:a, SN:b\\

BHNS:03& 1.3\ \%&
SCE:a$\rightarrow$b, SN:a, HCE:b$\rightarrow$a, NC:b$\rightarrow$a, SN:b\\

BHNS:04& 2.0\ \%&
all other\\

&&\\

BHBH:01& 17.7\ \%&
NC:a$\rightarrow$b, SN:a, HCE:b$\rightarrow$a, SN:b\\

BHBH:02& 10.5\ \%&
NC:a$\rightarrow$b, SCE:b$\rightarrow$a, SN:a, SN:b\\

BHBH:03& 1.4\ \%&
all other \\

\hline
\end{tabular}
\footnotesize{$^{\rm \alpha}$Normalized to the total DCO population.}\\
\footnotesize{$^{\rm \beta}$Sequences of different evolutionary phases for the primary (a) and the secondary (b): non-conservative MT (NC), single common envelope (SCE), double common envelope (DCE), common envelope with hyper-critical accretion (HCE), supernova explosion/core-collapse event (SN). Arrows mark direction of MT episodes.}

\normalsize

\noindent
in Table~\ref{channels}, the non-recycled NS-NS systems are formed via the
NSNS:09, NSNS:11 and NSNS:12 channels, which involve DCE of two low-mass
helium giants that were already allowed in the earlier version of {\em
StarTrack}.

The unique qualitative characteristic of this NS-NS formation path is
that both NS have avoided recycling. The NS progenitors have lost both their
hydrogen and helium envelopes prior to the two supernovae, so no accretion
from winds or Roche-lobe overflow is possible after NS formation.
Consequently, these systems are detectable as radio pulsars only for a
time ($\sim 10^6$\,yr) much shorter than recycled NS-NS pulsar lifetimes
($\sim 10^8-10^{10}$\,yr in the observed sample). Such short lifetimes are
of course consistent with the number of NS-NS binaries detected so far
and the absence of any {\em non-recycled} pulsars among them.

We note that the identification of the formation path for non-recycled
NS-NS binaries stems entirely from accounting for the evolution of helium
stars and for the possibility of double CE phases, both of which have
typically been ignored in previous calculations (with the exception of
\citealt{1998ApJ...496..333F}, where, however, such events were assumed to lead to 
mergers).

{\em Group II.} This subpopulation consists of tight, short lived
binaries with one recycled pulsar. Their merger times are typically
$\sim$\ 1 Myr or even smaller (see \S\,3.4.5). As shown in
Table~\ref{channels}, these new dominant NS-NS systems are formed via
the NSNS:01--08, NSNS:10 and NSNS:13 channels, with the common
characteristic that the {\em last} binary interaction is a
hyper-critical CE of a low-mass helium giant and the first-born NS.

In \citet{2002ApJ...571L.147B} we describe in detail the
formation of a typical NS-NS binary of group II. The  channel
identified as the most efficient for NS-NS formation (NSNS:01)
corresponds to the ``standard'' channel of \citet{1998ApJ...496..333F}.  The
only difference is an extra CE event which originates from allowing for
helium star evolution and without a priori assumptions about the CE
outcome. The second most dominant channel, involving two consecutive MT
episodes and then two SN explosions, closely resembles our channels:
NSNS:02, NSNS:04, NSNS:06, NSNS:10, NSNS:11, NSNS:12.  The only
difference again remains an extra MT episode from evolved,
Roche-lobe-filling helium stars.

The most dramatic effect of the binary evolution updates is reflected in
the existence of a whole new population of coalescing NS-NS stars formed
in Group II. In our standard model these channels contribute 50\% of
the DCO population, and their common characteristic is that the {\em
last} binary interaction is a hyper-critical CE of a low-mass helium
giant and the first-born NS\footnote{A more careful treatment of the response of helium stars to mass loss leads to a small reduction of this percentage, but the population forming through this channel is still significant, as discussed in \citet{2003ApJ...592..475I} and \citet{2006ApJ...648.1110B}}.  It turns out that the majority of these
systems survive the HCE event and form tight NS-NS binaries. Had we not
taken into account the radial expansion of low-mass helium-rich giants,
the progenitors of this dominant NS-NS population would have evolved
without any further MT. Most of them would have still formed NS-NS
systems, although not as tight as after this last CE episode. We have
actually examined this alternative and found that about half of them
would have formed binaries with merger lifetimes longer than 10$^{10}$
yr. 

{\em Group III.} This subpopulation consists of all the other NS-NS
systems  formed, through more or
less classical channels \citep{1991PhR...203....1B}. The
formation path denoted NSNS:14 corresponds to what is usually considered to be the ``standard'' NS-NS formation channel \citep{1991PhR...203....1B}. Since we account for hyper-critical accretion in CE, the formation rate is decreased because some NS (but not all, as assumed by \citealt{1998AA...332..173P} and by \citealt{1998ApJ...496..333F})  collapse
to BH. Furthermore our treatment of the hyper-critical accretion
typically leads to tighter post-CE systems, causing more binaries to
merge in CE events, and thus decreases the number of possible NS-NS
progenitors.

\citet{1995ApJ...440..270B} proposed a NS-NS formation channel where the progenitor consists of two massive stars with nearly equal masses (within 4\%) and the first mass transfer episode occurs when both stars are on the giant branch and have convective envelopes; if the binary survives the ensuing common envelope phase, both envelopes are ejected (double common envelope; DCE) and two helium stars are exposed which explode as supernovae and a NS-NS is formed. Brown was motivated by the almost equal mass measurements in NS-NS binaries and by adopting a low maximum NS mass (1.5\,M$_\odot$; associated with soft equations of state). In the proposed channel the progenitors are of almost equal mass and none of the NS ever experiences a common envelope phase, avoids collapsing into a BH. Some of the channels identified by \citet{2002ApJ...572..407B} (Table~\ref{channels}; specifically NSNS:09,11,12) are similar to the \citet{1995ApJ...440..270B} channel. However for the standard model they contribute only 8\% of the total NS-NS population because a more widely accepted maximum neutron star mass of 3\,M$_\odot$ is adopted and NS do not collapse into BH even if they experience hypercritical accretion in common envelopes (HCE). Recently Pinsonneault \& Stanek (2006) reported on a population of ``twin'' binaries with component masses within 5\%; this population about 50\% of the observed sample, but of course it is also favored by selection effects and the true contribution of ``twins'' could be closer to 25\% instead of 50\%. \citet{2002ApJ...572..407B} did also consider a model with a maximum NS mass of 1.5\,M$_\odot$ (model D2); as expected the NS-NS rate is decreased significantly especially in comparison to BH-NS, since most of the standard NS-NS channels in Table~\ref{channels} actually form BH-NS (the rate ratio shifts from 6.5 for the standard model A to 0.25 for model D2; see Table~4 in \citealt{2002ApJ...572..407B}). We note that a more recent population study of the Brown (1995) channel presented by Dewi et al.\ (2006) leads to NS-NS rates consistent with that from model D2.

\subsection{Populations of Black Hole Binaries} 

In general, BH-NS and BH-BH binaries are formed through just a few
distinct channels, with a moderate number of MT events (2--3), in
contrast to our findings for NS-NS systems.
Helium star evolution, radial expansion and CE phases are much less
important for their formation.  The reason is
that for most of these progenitors the first-born compact object is
massive enough that they do not expand to large radii nor they lead to possible CE evolution
(see  channel BHNS:03 in
Table~\ref{channels}).  Instead these DCO form most efficiently through
channels that closely resemble those NS-NS conventionally thought to be ``standard'' \citep{1991PhR...203....1B, 1998ApJ...496..333F}: evolution is initiated with a phase of non-conservative mass transfer and followed either by a CE phase or the formation of the first compact object (see BHNS:01, BHNS:02, and BHBH:01, BHBH:02).

\section{Merger Rates of Double Neutron Stars: Observed Pulsar Sample}
\label{empirical} 

\subsection{Introduction}

Double Neutron Stars that will merge within a Hubble time are one of the prime targets for ground-based gravitational-wave (GW) interferometers such as GEO600, TAMA, VIRGO, and LIGO. Event rates of the DNS inspiral searches by these detectors can be inferred using the rate estimates with an extrapolation out to the maximum detection distances for any detector under consideration. Before 2003, the Galactic DNS merger rate had been estimated between $\sim10^{-7} - 10^{-5}$ yr$^{-1}$ (see \citealt{2001ApJ...556..340K} and references therein). At that time, there were only two systems available for empirical studies, PSRs~B1913+16 \citep{1975ApJ...195L..51H} and B1534+12 \citep{1991Natur.350..688W}. We calculated the PDF of Galactic DNS merger rates, $P(R)$, based on these two systems (see \citealt{2003ApJ...584..985K}, hereafter KKL, for further discussion\footnote{We note that a web interface that allows the calculation of $P(R)$ for a wide range of pulsar surveys and pulsars is publicly available at http://www.astro.northwestern.edu/$\sim$ciel/gppg\_main.html}). Soon after the discovery of the highly relativistic system PSR~J0737$-$3039 \nocite{2003Natur.426..531B}, we were able to revise $P(R)$ including PSR~J0737$-$3039 in collaboration with the observation team\footnote{Only the millisecond component (J0737-3039A) is considered in our calculation. For instance, the current age of the system is derived from pulsar A.}\citep{2003Natur.426..531B, 2004ApJ...601L.179K}. The discovery of J0737$-$3039, resulted in a significant  increase in the estimated Galactic DNS merger rate by a factor of $\sim$6. This implies a boost in event rates for DNS searches for GW interferometers. Here, we present our recent results on: (i) the PDF of Galactic DNS merger rate estimates with updated observations; (ii) the {\em global\/} PDF of rate estimates considering the systematic uncertainties; (iii) constraints on upper limits for rate estimates based on the observed supernova rate incorporated with our theoretical understanding of the SN-DNS relation; (iv) the approximate contribution of J1756$-$2251 to the Galactic DNS merger rates and uncertainties in our assumptions on the efficiency of the acceleration searches for Parkes multibeam pulsar survey (PMPS). Finally, we discuss implications of the most recently discovered pulsar binary J1906+0746 \citep{2006ApJ...640..428L}.

\subsection{The Galactic DNS merger rate}
Here, we describe the main components of the calculation of the combined $P({\Rate})$ considering the three observed DNS systems in the Galactic disk. The details of these calculations are discussed by KKL and \citep{2004ApJ...601L.179K}. The merger rate of a given population {\em i} can be defined by 

\begin{equation}
R_{i} \equiv \left(\frac{N_{\rm PSR}}{\tau_{\rm life}}f_{\rm b} \right)_{i} ~,
\end{equation}

where $N_{\rm PSR, i}$ represents the number of pulsars in our Galaxy with pulse and orbital characteristics {\it similar\/} to an observed sample {\em i} (e.g. PSR J0739$-$3039) and $f_{\rm b, i}$ is a correction factor to take into account pulsar beaming (typically $\sim 6$)\footnote{This is based on polarimetry measurements of B1913+16 and B1534+12.}; $\tau_{\rm life, i}$ is the lifetime of system {\em i} based on its observed properties. In order to calculate $N_{\rm PSR, i}$, we distribute a large population of pulsars {\em all}~similar to the system {\em i}, in a model galaxy assuming spatial and luminosity distributions. Since pulsar luminosities are drawn from a distribution, the observed flux and estimated distance for the DNS system are not relevant in our calculation. Moreover, the selection effects for faint pulsars are implicitly taken into account. Once we generate a pulsar population with a size $N_{\rm PSR}$, we can then calculate the number of pulsars detected ($N_{\rm det}$) by large-scale pulsar surveys. We repeat the survey simulations with a detailed modeling of selection effects for observed DNS systems. For a fixed $N_{\rm PSR}$, we find that $N_{\rm det}$ follows a Poisson distribution, P($N_{\rm det}; \bar N_{\rm det}$), where $\bar N_{\rm det}$ is a mean value of $N_{\rm det}$ for a given population size ($N_{\rm PSR}$). We require  $N_{\rm det}=1$, i.e., we consider only one observed system, and calculate the best-fit value of $\bar N_{\rm det}$. With a wide range of $N_{\rm PSR}$, we find $\bar N_{\rm det} = cN_{\rm PSR}$ as expected where c is a constant. Applying Bayes' theorem to these results, we calculate a P($N_{\rm PSR}$), a PDF for the population size of a given system {\em i} knowing that there is one observation. The calculation $P(R)$ from $P(N_{\rm PSR})$ is straightforwad, and the full derivation and the analytical formula can be found in Appendix~A of \citet{2004ApJ...616.1109K}.

The lifetime of a DNS system is defined by $\tau_{\rm life} \equiv
\tau_{\rm sd} + \tau_{\rm mrg}$, where $\tau_{\rm sd}$ is the spin-down
age of a recycled pulsar \citep{1999ApJ...520..696A} and $\tau_{\rm mrg}$ is the remaining lifetime until the two neutron stars merge \citep{pm63}.
Based on the most recent observations, we estimate the lifetime of J0737$-$3039 to be $\sim$230\,Myr \citep{2004Sci...303.1153L}. This is the shortest among the three observed systems. The beaming correction factor $f_{\rm b}$ is defined as the inverse of the fractional solid angle subtended by the pulsar beam. Its
calculation requires detailed geometrical information on the
beam. Following \citet{2001ApJ...556..340K} and updating the values with recent observations, 
we calcualte $f_{\rm b}$$\sim$5.7 for PSR B1913+16 \citep{2002ApJ...576..942W} and $\sim$6.0 for PSR B1534+12 \citep{2004PhRvL..93n1101S} Without good knowledge of the geometry of J0737$-$3039, we assume $f_{\rm b, J0737}$ to be the average value of the other two systems ($\simeq 5.9$).

\begin{figure}
\begin{center}
\includegraphics*[width=4.in]{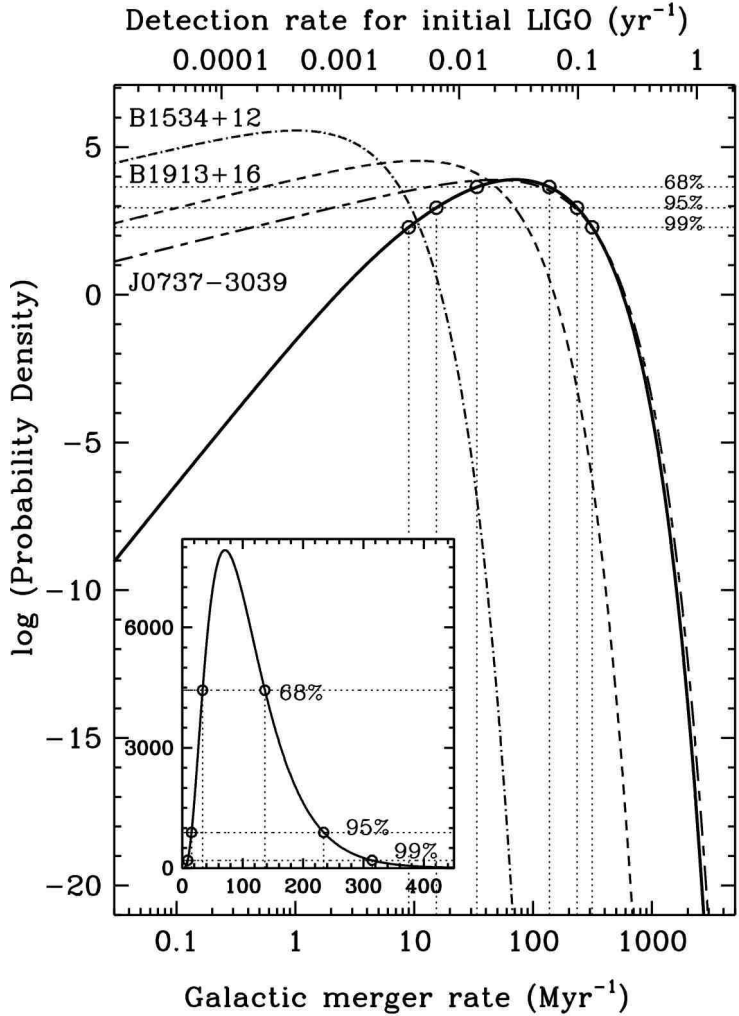}
\end{center}
\vspace{.6cm}
\caption{$P({\Rate})$ is shown on a log
scale. The thick solid line is the Galactic rate estimate
overlapped with results for individual observed systems (dashed
lines). Dotted lines indicate confidence intervals for the rate
estimates. The same results are shown on a linear scale in the small
inset. All results are from our reference model.}
\end{figure}

In Fig.\ 9, we show $P({\Rate})$ for a reference model (Model~6 in KKL\nocite{2003ApJ...584..985K}). We obtain the most likely value of ${\Rate} \sim 71\,{\rm Myr}^{-1}$,
larger by a factor of $\simeq 5.5$ than the rate estimated before the
discovery of J0737$-$3039. The increase factor is found similar for all
pulsar population models we examined. The increased merger rates imply a boost in the inferred detection rate of GWs from DNS inspirals for ground-based GW interferometers such as LIGO. In order to calculate the detection rate ($D$), we assume a homogeneous distribution of galaxies in nearby Universe and a spherical symmetry in detector sensitivity. Then, we can write ${\rm D} \equiv R_{\rm gal} \times N_{\rm gal}$, where $N_{\rm gal}$ is the number of galaxies in the detection volumne ($V_{\rm det}$). We calculate the number density of galaxies derived by the observed blue luminosity density, ($n_{\rm gal} = 1.25\times 10^{-2}$Myr$^{\rm -1}$ (see \citet{1991ApJ...380L..17P} and \citet{2001ApJ...556..340K} for more details). The detection volume of LIGO can then be defined as a sphere for a given detection distance (20 Mpc and 350 Mpc for the inital and advanced LIGO, respectively), and the number of galaxies within $V_{\rm det}$ is simply $n_{\rm gal} \times V_{\rm det}$.  For our reference model, we find that the most probable event rates are about 1 per 30\,yrs and 1 per 2 days, for initial and advanced LIGO, respectively. At the 95\% confidence interval, the most optimistic predictions for the reference model are 1 event per 9\,yrs and 1.6 events per day for initial and advanced LIGO, respectively. More details can be found in \citet{2004ApJ...601L.179K}.

As shown in Fig.\ 9, the Galactic DNS merger rate is dominated by PSR J0737$-$3039. We note that the current age of 30-70Myr for J0737$-$3039, suggested 
by \citet{2005ASPC..328..113L}, implies a even shorter lifetime
 ($\tau_{\rm life} \sim115-155$ Myr) knowing that the estimated merger timescale of this system is $\sim$85 Myr. Based on their results, we find the most likely value of $\Rate$ for the reference model is $\simeq 90-110\,$Myr$^{-1}$.

The beaming correction for J0737$-$3039 is not yet constrained and we assume MSPs discovered in DNS systems are not very different. As a conservative lower limit, without any beaming corrections for all observed systems for a reference model, we obtain $\sim 12^{+11}_{-6}$ Myr$^{-1}$ with a 95\% confidence interval. The corresponding detection rates for initial and advanced LIGO are $5^{+12}_{-4} \times 10^{-3}$ yr$^{-1}$ and $27^{+62}_{-21}$ yr$^{-1}$, respectively. Only when the axes geometry of J0737-3039 becomes available, we will be able to constrain the uncertainties of the beaming fraction, and in turn, the rate estimates. 

\subsection{Global probability distribution of the rate estimates}

In KKL, we showed that empirical DNS merger rates are strongly dependent on the assumed luminosity distribution function for pulsars, but not on the pulsar spatial distribution. Therefore, we can consider only the rate dependence on the pulsar luminosity function for simplicity. Here, we describe how we can incorporate the systematic uncertainties from these models and calculate, $P_{\rm g} ({\Rate})$, a {\em global\/} PDF of rate estimates. Note that the results available on prior functions for pulsar luminosity distribution are currently out of date. Specific quantitative results could change when constraints on the luminosity function are derived from the current pulsar sample.

We assume a power-law luminosity distribution for a radio pulsar luminosity function $f(L)$. This function is defined by two parameters: the cut-off luminosity $L_{\rm min}$ and power-index $p$. We assume prior distributions for these two parameters and calculate $P_{\rm g}({\Rate})$. We fit the marginalised likelihood of $L_{\rm min}$ and $p$ presented by \citet{1997ApJ...482..971C} and obtain the following analytic formulae for prior functions, i.e. $f(L_{\rm min})$ and $g(p)$: $f(L_{\rm min}) = \alpha_{\rm 0} + \alpha_{\rm 1} L_{\rm min} + \alpha_{\rm 2} L_{\rm min}^{2}$ and $g(p) = 10^{\beta_{\rm 0} + \beta_{\rm 1} p + \beta_{\rm 2} p^{2}}$, where $\alpha_{\rm i}$ and $\beta_{\rm i} ~({\rm i}=0,1,2)$ are coefficients we obtain from the least-square fits and the functions are defined over the intervals $L_{\rm min}=[0.0,\,1.7]$ mJy kpc$^{2}$ and $p=[1.4,\, 2.6]$. We note that, although \citet{1997ApJ...482..971C} obtained $f(L_{\rm min})$ over $L_{\rm min}\simeq [0.3,\,2]$ mJy kpc$^{2}$ centered at 1.1 mJy kpc$^{2}$, we consider $f(L_{\rm min})$ with a peak at $\sim 0.8\,$mJy kpc$^{2}$ considering the discoveries of faint pulsars with L$_{\rm 1400}$ below 1 mJy kpc$^{2}$ \citep{2003ASPC..302..145C}. Next we calculate $P_{\rm g}({\Rate})$ as follows: 
\begin{equation}
P_{\rm g}({\Rate})
= \int_{p}dp \int_{L_{\rm min}}dL_{\rm min} P(R)f(L_{\rm min})g(p).
\end{equation}

\begin{figure}
\begin{center}
\includegraphics*[width=4.in]{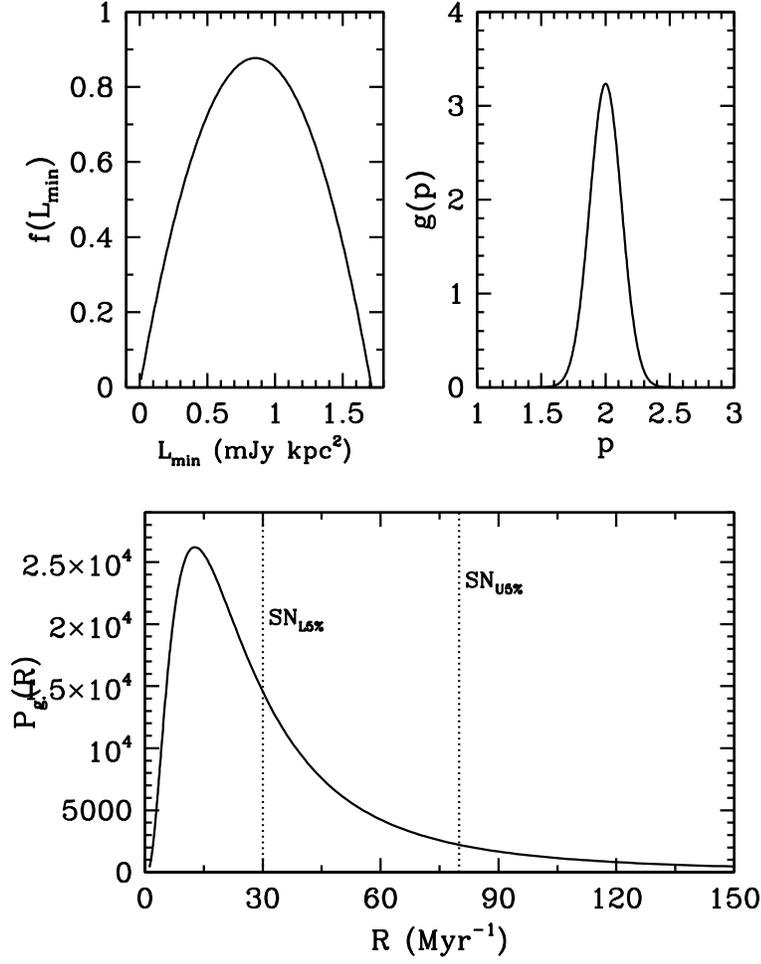}
\end{center}
\vspace{.6cm}
\caption{The global $P_{\rm g}({\Rate})$ on a linear scale (lower panel) and the assumed intrinsic distributions for $L_{\rm min}$ and $p$ (upper panels). Dotted lines represent the lower ($SN_{\rm L}$) and upper ($SN_{\rm U}$) bounds on the observed SN Ib/c rate scaled by 5\% of the observed SN Ib/c rates, $600-1600$ Myr$^{-1}$ (see text).}
\end{figure}

In Fig.\ 10, we show the distributions of $L_{\rm min}$ and $p$ adopted (top panels) and the resulting global distribution of Galactic DNS merger rate estimates (bottom panel). We find the peak value of $P_{\rm g}({\Rate})$ at {\em only\/} around 13\,Myr$^{-1}$. We note that this is a factor $\sim 5.5$ smaller than the result from our reference model (${\Rate}\sim$71\,Myr$^{-1}$). At the 95\% confidence interval, we find that the {\it global} Galactic DNS merger rates lie in the range $\sim$ 1--145\,Myr$^{-1}$. These imply LIGO event rates in the range $\sim (0.4-60)\times 10^{-3}\,$yr$^{-1}$ (initial) and $\sim 2-330\,$yr$^{-1}$ (advanced). Since 1997, the number of known millisecond pulsars has more than doubled, and therefore, constraints on $L_{\rm min}$ and $p$ and their PDFs based on the most up-to-date pulsar sample are urgently needed.

\subsection{Upper Limit of DNS Merger Rate Estimates. Constraints from Type Ib/c Supernovae Rates}
According to the standard binary evolution scenario, 
the progenitor of the second neutron star is expected to form during a Type Ib/c supernova. 
Therefore, the empirical estimates for the Type Ib/c SN rate in our Galaxy can be used to set upper limits on the DNS merger rate estimates. Based on \citet{1999A&A...351..459C}, we adopt ${\Rate}_{\rm SN\,Ib/c} \simeq 1100\pm500\,$Myr$^{-1}$ considering Sbc--Sd galaxies, a Hubble constant $H_0=71\,$km/s/Mpc and the blue luminosity of our Galaxy $L_{\rm B,gal}=9\times10^{9}\,L_{\rm sun}$.

We calculate the fraction of SN Ib/c actually involved in the formation of DNS with a binary evolution code {\em StarTrack} \citep{2002ApJ...572..407B, 2005astro.ph.11811B} and estimate the rate ratio: $\gamma\equiv({\Rate}_{\rm DNS}$/${\Rate}_{\rm SN\,Ib/c})\times100\le5\%$. 
Motivated by this result, we adopt the empirical ${\Rate}_{\rm SN\,Ib/c}$ assuming $\gamma\sim5\%$ and compare the value with the global PDF (Fig. 10).

We note that our most optimistic DNS merger rate is ${\Rate}=189^{+691}_{-166}$ Myr$^{-1}$ at a 95\% confidence interval (Model~15 in KKL\nocite{2003ApJ...584..985K}). We obtain $\gamma \sim 80\%$ with respect to the center value of the empirical SN Type Ib/c rate ($1100$ Myr$^{-1}$) and the upper limit of ${\Rate}$ at the 95\% confidence interval ($189+691=880$ Myr$^{-1}$). This corresponds to $\gamma \sim 13\%$ for a SN Type II rate, which is factor 6.1 larger than that of SN Type Ib/c. In both cases, the most optimistic model is lower than the current empirical supernova rate estimates, but not really consistent with the results of population synthesis calculations. If we consider an upper limit of ${\Rate}$ at the 95\% confidence interval from the global PDF, we obtain $\gamma\sim$ 13\% and 2\% for the center value of SN Type Ib/c (1100 Myr$^{-1}$) and II, respectively.

\subsection{Implications of New Discoveries to the Galactic DNS merger rate estimates}

Recently, \citet{2005ApJ...618L.119F} discovered PSR J1756$-$2251, the 4th merging DNS in the Galactic disk from the Parkes Multibeam Pulsar Survey (PMPS). The standard Fourier method failed to find this pulsar and they reanalysed the PMPS data with an acceleration search (or a `stack search' as described in their paper). In order to calculate the merger rate including PSR J1756$-$2251, a detailed simulation is necessary to calculate the effect of the acceleration search with the PMPS. However, the approximate contribution of PSR J1756$-$2251 to the Galactic merger rate can be easily obtained. We find the total rate increases by only $\sim4\%$ due to the new discovery. This is expected because PSR J1756$-$2251 can be identified as a member of the B1913+16-like population, which has already been taken into account in the calculation. Only future detections of pulsars from a significantly different population (compared to the known systems), or from the most relativistic systems, will result in a non-trivial contribution to the rate estimates. 

Finally, we note the implications of J1906+0746 on the pulsar binary merger rates. This system has drawn attention due to the extremely young age of the pulsar (characteristic age of $\sim$ 112 kyr; \citealt{2006ApJ...640..428L}). If the companion is another neutron star, J1906+0746 would be the first discovery of a non-recycled component in a DNS system. Currently, the nature of the companion is totally unknown, and it can be either a light neutron star, or heavy (O-Mg-Ne) white dwarf. Assuming J1906+0746 is a DNS system, we calculate its contribution to the Galactic merger rate. Because of its short lifetime ($\sim$82Myr), J1906+0746 can increase the Galactic DNS merger rate by about a factor 2. This implies that the current estimated DNS merger rate including J0737$-$3039 can still be doubled! If J1906+0746 is an eccentric NS--WD system, such as J1141-6545 \citep{1999ApJ...516L..25V} or B2303+46 \citep{1985ApJ...294L..21S}, it will be as important as J1141-6545, which is currently dominate the birthrate of eccentric NS-WD binaries.

\section{Merger Rates of Double Compact Objects: Population Synthesis}
\label{popsyn} 

Lacking a sample of  DCOs containing a black hole,
the only route to BH-BH merger rates is via population synthesis models.
These involve a Monte Carlo exploration of the likely
life histories of binary stars, given 
statistics governing the initial conditions for binaries and 
a method for following the behavior of single and
binary stars 
\citep[see, e.g.,][and references therein]{2002ApJ...572..407B, 2002MNRAS.329..897H, 1996AA...309..179P, 1998ApJ...496..333F}.
Unfortunately, our understanding of the evolution of single and binary
stars is incomplete, and we parameterize the associated uncertainties with a great
many parameters ($\sim 30$), many of which can cause the predicted
DCO merger rates to vary by more than one or two orders of
magnitude when varied independently through their plausible range.
To arrive at more definitive answers for DCO merger rates,
we must substantially reduce our uncertainty in the parameters that
enter into population synthesis calculations through comparison with
observations when possible.

The simplest and most direct way to constrain the parameters
of a given population synthesis code is to compare several of  its
many predictions against observations.
For example, \citet{2005ApJ...633.1076O} required their population synthesis
models be consistent with the  empirically
estimated formation rates derived from  
the four known Galactic \nsns binaries  which are tight
enough to merge through the emission of gravitational waves within
$10$\,Gyr.  
Here, we require our population synthesis models to be
consistent (modulo experimental error) with six observationally
determined rates: (i) the formation rate 
implied by the  known Galactic merging \nsns binaries, mentioned
above; (ii) the formation rate implied by the known Galactic
\nsns binaries which do \emph{not} merge within $10$ Gyr (henceforth denoted ``wide''
\nsns binaries); (iii,iv) the formation rate  implied by the sample of
merging and eccentric WD-NS binaries; and finally (v,vi) the type II and
type Ib/c SNe rates.
Further, we use the set of models consistent with these constraints to
revise our population-synthesis-based expectations for various DCO
merger rates, assuming no prior information, so all 
population synthesis model parameters consistent with our constraints
are treated equally.

\subsection{Observations of DCOs}
\label{sec:data:DCOs}
Seven  \nsns binaries and four WD-NS binaries (with relatively massive WDs) have been
discovered so far in the galactic disk, using pulsar surveys with
well-understood selection effects.  [We are very specifically
\emph{not} including the recently-discovered binary PSR J1906+0746
\citep{2006ApJ...640..428L}, because the companion cannot be
definitively classified as a WD or NS at present. We also omit
PSR J2127+11C found in the globular cluster M15, since its formation is thought to be dynamical and not due to isolated binary evolution in the Galactic field.

\subsection{Methodology I: Preferred population model}

As already discussed, in the context of \nsns  KKL developed a statistical method to
estimate the likelihood 
of DCO formation rates, given an observed sample of DCOs in which one
member is a pulsar, designed to account
for the small number of known systems and their associated
uncertainties.  
In this section, we present  results only for our
reference pulsar luminosity distribution model,  corresponding a 
power law luminosity distribution 
with negative slope, index $p=2$, and minimum luminosity $L_{min}=0.3$~mJy kpc$^2$ 
[model 6 of KKL; as discussed therein, this model
better accounts for more recent observations of faint pulsars]; in the
following section, we describe our our predictions change when
systematic uncertainties in $p$ and $L_{min}$ are incorporated into
${\cal P}$.

Finally, for each class of binary pulsars we define symmetric 95\% confidence intervals:
 the upper and lower rate limits ${\cal R}_{w,\pm}$ satisfy
\begin{equation}
\int_0^{{\cal R}_{w,-}} d {\cal R} {\cal P}_w({\cal R}) =
\int_{{\cal R}_{w,+}}^\infty d {\cal R} {\cal P}_w({\cal R}) = 0.025
\; .
\end{equation}
This confidence-interval convention is different from with the customary
choice presented in KKL.

In all plots that follow, rate probability distributions are
represented using a logarithmic scale for ${\cal R}$; thus, instead of
plotting ${\cal P}$, all plots instead show
\[
p(\log {\cal R}) = {\cal P}({\cal R}) {\cal R} \ln 10 \; .
\]

\subsubsection{WD-NS binaries}
\label{sec:dco:wdns}
Four WD-NS binaries with relatively massive WDs have been discovered in the galactic disk: PSRs
J0751+1807, J1757-5322, J1141-6545, and B2303+46.   While all four
binaries may be applied to a \emph{net} WD-NS formation rate estimate,
the sample manifestly contains relics of distinctly different
evolutionary channels; for example, while J0751+1807 and J1757-5332
have evidently been strongly circularized and spun up by a recent mass transfer
episode, J1141-6545 cannot have been  \citep{2000ApJ...543..321K, 2003ApJ...595L..49B}.  
Ideally, population synthesis must produce distributions of WD-NS
binaries consistent with both the observed orbital parameters and
spins; however, the present sparse sample does not allow a reliable
nonparametric estimate of the distribution of WD-NS binary
parameters.  Instead, as first step towards applying constraints based
on binary parameter \emph{distributions}, we
subdivide these five binaries into two overlapping classes:
\emph{merging} binaries, denoted WD-NS(m), which will merge through the
emission of gravitational waves within $10$ Gyr; and  \emph{eccentric}
binaries, denoted WD-NS(e), which have significant ($e>0.1$)
eccentricity at present.  The rate estimate derived for  both classes is
dominated by J1141-6545  \citep{2004ApJ...616.1109K, 2005ASPC..328..261K}.

\subsubsection{\nsns\ binaries}
\label{sec:dco:nsns}
Seven \nsns binaries have been discovered so far in the Galactic disk.
Four of the known systems will have merged within $10$\,Gyr (i.e.,
``merging'' binaries: PSRs J0737-3039, B1913+16, B1534+12, and
J1756-2251) and three are wide with much longer merger times (PSRs
J1811-1736, J1518+4904, and J1829+2456). PSR~J1756-2251 was discovered
recently with acceleration searches \citep{2005ApJ...618L.119F}.  
As already discussed, the rate increase is
estimated to be smaller than 4\%.   For this reason, we omit it when
estimating NS-NS merger rates.

The observed NS-NS population naturally subdivides into two distinct classes, 
 depending on whether
they merge due to the emission of gravitational waves within 10 Gyr:
\emph{merging} \nsns binaries, denoted NSNS(m), and \emph{wide}
\nsns binaries, denoted NSNS(vw).   

\subsubsection{Recycling, selection effects,  and the lack of wide NS-NS
  binaries}
\label{sec:dcos:recycling}
We find that the confidence
intervals for  wide and merging NS-NS binaries are almost an order of
magnitude from overlapping.  On the contrary, 
population
synthesis simulations produce merging and wide binaries at a roughly
equal (and always highly correlated) rates.  Since our rate estimation
technique automatically compensates for the most obvious selection
effects (e.g., orientation and detectable lifetime),  two
unbiased samples of wide and merging NS-NS binaries should arrive at
nearly the \emph{same} prediction for the NS-NS formation rate.

\citet{2005ApJ...633.1076O}
explained this significant discrepancy by a selection effect:
evolutionary tracks leading to wide NS-NS binaries should be less likely to
recycle the first-born pulsar.  The \emph{observed} sample of wide
NS-NS binaries [NSNS(vw)] represents a much smaller subset of
\emph{recycled} pulsars.   As summarized in \citet{PSutil2},
we assume any wide NS-NS binary which in its past underwent
\emph{any conventional} mass transfer  will recycle one of its neutron
stars to a pulsar. This  condition is likely to over-estimate  the true
NSNS(vw) formation rate. We note that neutron-star recycling can also possibly happen during dynamically unstable mass-transfer phases (common envelopes), but the vast majority of wide NS-NS do not evolve through such a phase.

\subsubsection{Pulsar Population Model Uncertainties}
As noted in KKL and subsequent papers, our reconstruction of the
pulsar population (i.e., $N_{PSR}$) relies upon our understanding of
pulsar survey selection effects and thus on the underlying pulsar luminosity
distribution.  This distribution can be well-constrained
experimentally \citep[see, e.g.][]{1997ApJ...482..971C}, though these
constraints do not yet incorporate  recent
faint pulsar discoveries such as J1124-5916
\citep{2002ApJ...567L..71C}. 
Nonetheless, different observationally-consistent
distributions imply significantly different distributions, with
maximum-likelihood rates differing by factors of order $10$
\citep{2003ApJ...584..985K}. 
Since the constraint intervals discussed above assume the \emph{preferred pulsar
  luminosity distribution model} --  a power-law pulsar luminosity
distribution with negative slope $p=2$ and minimum cutoff luminosity
$L_{min}=0.3$ mJy kpc$^2$
-- they do not incorporate any uncertainty in the pulsar luminosity function.

The infrastructure needed to incorporate uncertainties in the pulsar
luminosity function has been presented in \S\,4.2. 
However, out-of-date constraints on the pulsar luminosity function
allow implausibly high minimum
pulsar luminosities $L_{min}$.  A high minimum pulsar luminosity
implies fewer merging pulsars have been missed by surveys.  Thus, these
out-of-date constraints on pulsar luminosity functions permit models
consistent with 
substantially lower merger rates than now seem likely, given the
discovery of faint pulsars.  In other words, if we use the infrastructure
presented in \citet{2006.astro-ph..0608280} to  marginalize over $L_{min}$
and $p$ generate a \emph{net} distribution function for the merger
rate, then the 95\% confidence intervals associated with that net
distribution would have a spuriously small lower bound, entirely
because the pulsar population model permits large $L_{min}$.
Therefore, we present results based only on our
preferred luminosity function and do not include the out-of-date
luminosity function constraints by \cite{1997ApJ...482..971C} and the
related net rate distribution provided by \citet{2006.astro-ph..0608280}.

\subsection{Observations of supernovae}
\label{sec:data:other}

Type Ib/c and II supernovae occur extremely rarely near the Milky Way.  
While historical data contains several observations of and even
surveys for supernovae, the selection effects in these long-duration
heterogeneous data sets have made their interpretation difficult
(Cappellaro 2005, private communication).  In this paper, we estimate
supernova rates and uncertainties via Table 4 of
\citet{1999A&A...351..459C}.  \citet{1999A&A...351..459C} present their results
in terms of  a number of supernovae per century per $10^{10}$ blue solar
luminosities ($L_{\odot,B}$); to convert their results to rates per Milky Way
equivalent galaxy, we 
assume a Milky Way blue luminosity of 
$2\times 10^{10}L_{\odot,B}=0.9\times 10^{10} L_{\odot}$ relative to
the solar blue $(L_{\odot,B})$ and total $(L_\odot)$ luminosities
(see, e.g., 
\citet{2001ApJ...556..340K},
\citet{1991ApJ...380L..17P}, \citet{2000asqu.book.....C}
and references therein).
Using their estimates for the most-likely supernovae
rates and for the $1\sigma$ errors in those rates for Milky Way-like
galaxies (i.e., S0a-Sb), we arrive at the $2\sigma$
logarithmic confidence intervals used in our constraints. 

Though fairly accurate studies exist of the high-redshift
supernova rate \citep[e.g.,][]{2005A&A...430...83C}, they have little  relevance
to the present-day Milky Way.
Several surveys
have also attempted to determine the supernova rate in the Milky Way by a variety of
indirect methods, such as statistics of supernova remnants
\citep[highly unreliable due to  challenging selection effects; see, e.g.,][]{1991ARAA..29..363V}
and direct observation of radioisotope-produced backgrounds \citep[e.g., decay from ${}^{26}Al$, as described
in][]{2006Natur.439...45D}.  Taken independently, these methods
have greater uncertainties than the historical studies of
\citet{1999A&A...351..459C}.

\subsection{Population synthesis predictions}
\label{sec:ps}

\subsubsection{{\em StarTrack} population synthesis code}

We estimate formation and merger rates for several classes of double
compact objects using the 
\emph{StarTrack} code first developed by \citet{2002ApJ...572..407B} and recently significantly updated and tested as described in detail in
\citet{2005astro.ph.11811B}.   
This updated code predicts somewhat different double compact object properties than the version used in \citet{2002ApJ...572..407B};  a forthcoming paper (Belczynski et al., in preparation) will discuss these changes and the evolutionary physics underlying them in significantly greater detail.

Like any population synthesis code, it evolves randomly chosen
binaries from their birth to the present, tracking the stellar and
binary parameters.  For any  class of events that is
\emph{identifiable within the code}, such as supernovae or DCO
mergers, we estimate event rates by taking the average event rate
within the simulation (i.e., by dividing the
total number of events seen within some simulation by the duration of
that simulation) and renormalizing by a scale factor that depends on
properties of the simulation (i.e., the number of binaries simulated
and the binary birth mass distributions assumed) and the Milky Way as
a whole (i.e., the present-day star formation rate); see Eq.~(2) and
Appendix A of \citet{2005ApJ...620..385O} for details. Specifically, our simulations\footnote{Our approach gives only the \emph{average} event rate.  The
  present-day merger rate agrees with this quantity when most mergers occur
  relatively promptly (e.g., $<100$ Myr) after their birth.
  Some DCOs -- notably double BH binaries --  have substantial delays
  between birth and merger, introducing a strong time dependence to
  the merger rate.  The technique described above will significantly
  \emph{underestimate} these rates.  This point will be addressed in
  considerably more detail, both for the Milky Way and for a
  heterogeneous galaxy population in a forthcoming paper by
  \citet{PSutil2}.
} are normalized to be consistent with an assumed Milky Way star formation rate of $3.5 M_\odot {\rm yr}^{-1}$.

When constructing our archived population synthesis results, we did
not choose to record detailed information about the nature and amount
of any mass transfer onto the first-born NS.  We therefore cannot
reconstruct precise population synthesis predictions for the NSNS(vw)
formation rate.  However, we do record whether some mass transfer
occurs, and the nature of the mass transfer mechanism.  The
conventional mass transfer mechanism -- dynamically stable Roche-lobe
overflow -- inevitably produces a disk around the compact  object and
can potentially spin that object up.   Other mechanisms, such as
(possibly hypercritical) common-envelope evolution, presumably involve
a substantially more spherical accretion flow; the specific angular
momentum accreted may be substantially lower, possibly not enough to
recycle the neutron star. Thus for the purposes of
identifying a class of potentially recycled (``visible'') wide NS-NS
binaries, we assume any system which underwent mass transfer  (dynamically stable in the case of wide systems) produced a recycled NS primary. 

\begin{figure}
\begin{center}
\includegraphics[width=5.0in]{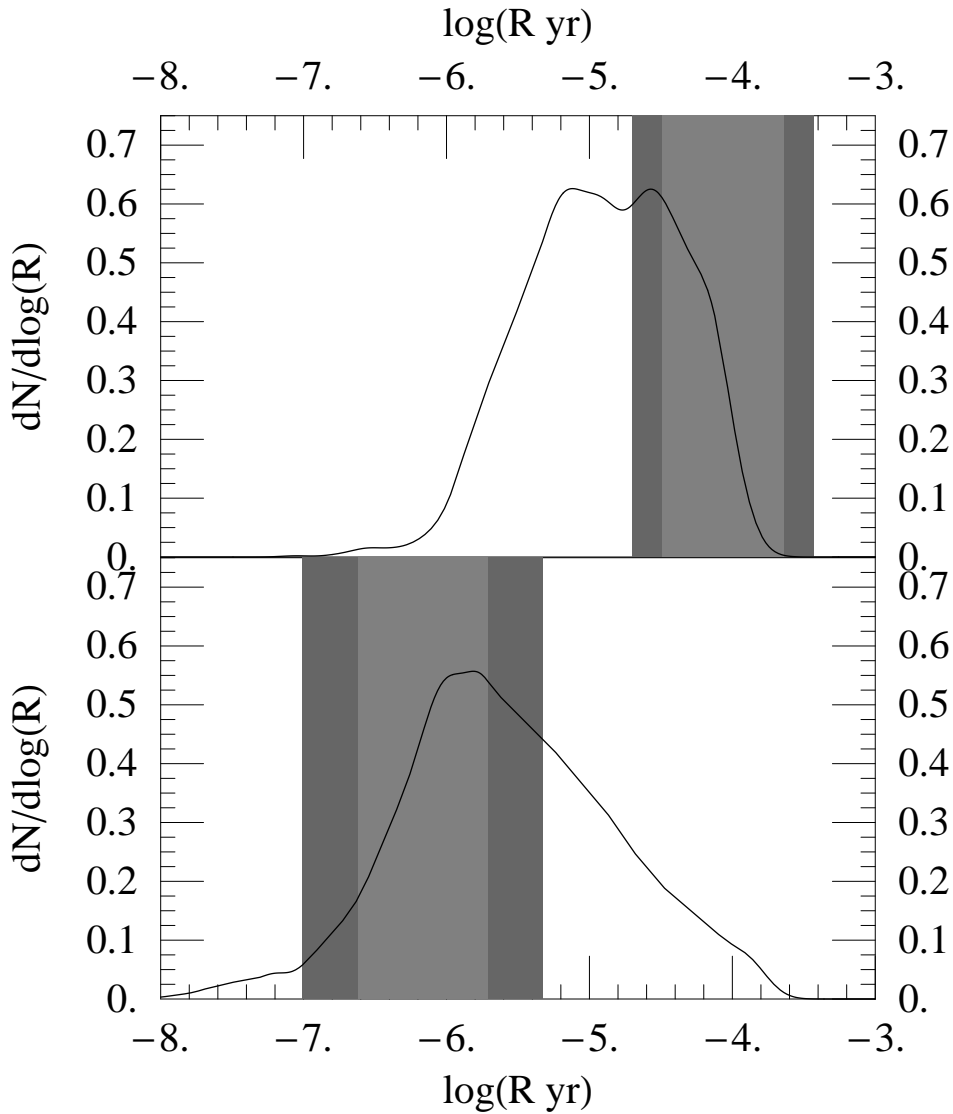}
\end{center}
\caption{\label{fig:ConstraintsVersusDistribs:NSNS} 
Probability
  distributions for merging (top) and wide (bottom) NS-NS
  formation rates, as predicted by
  population synthesis.  The light shaded region shows the 95\% confidence interval
  consistent with observations of binary pulsars, as described in
  Sec.~\ref{sec:data:DCOs}.
Because of sparse sampling and
  relatively poor data, our fit for the
  formation rate of visible, wide NS-NS binaries is significantly more
  uncertain than other fits.
}
\end{figure}

\begin{figure}
\begin{center}
\includegraphics[width=5.in]{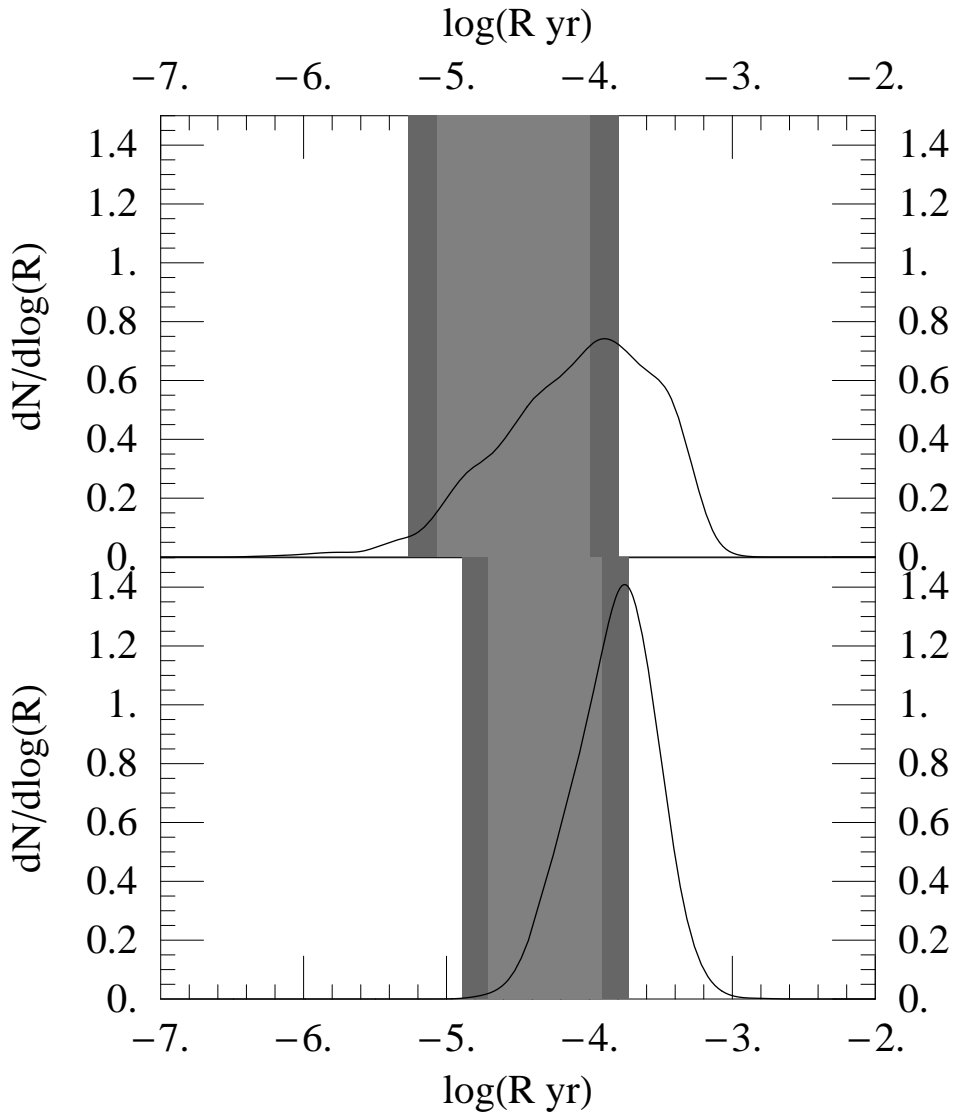}
\end{center}
\caption{\label{fig:ConstraintsVersusDistribs:WDNS} Probability
  distributions for merging (top) and eccentric (bottom) WD-NS
  formation rates, as predicted by
  population synthesis.  The light shaded region shows the 95\% confidence interval
  consistent with observations of binary pulsars, as described in
  Sec.~\ref{sec:data:DCOs}.  The dark shaded region extends this
  constraint interval by 
an estimate of the
  systematic error in each fit; see \citet{PSutil2}.    
}
\end{figure}

\begin{figure}
\begin{center}
 \includegraphics[width=5.in]{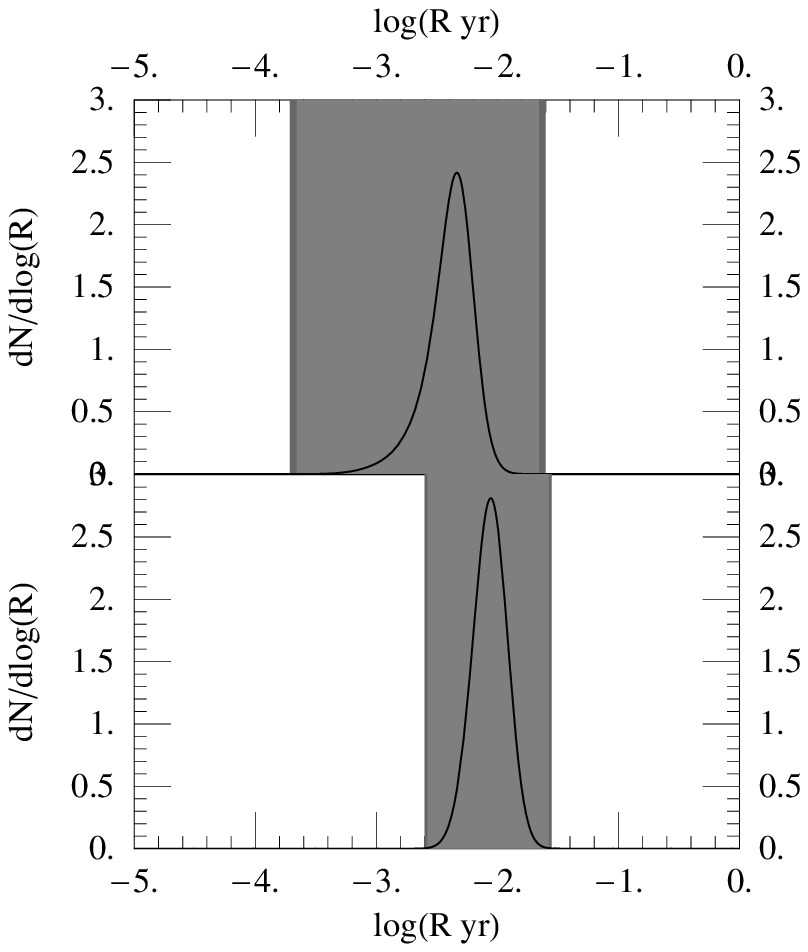}
 \end{center}
\caption{\label{fig:ConstraintsVersusDistribs:SN} Probability
  distributions for SN Ib/c (top) and SN II (bottom), as predicted by
  population syntheses with \emph{StarTrack}.  The light shaded region
  shows the ($\approx 2\sigma$) interval
  consistent with observational constraints of supernovae, from
  \citet{1999A&A...351..459C}.  A dark shaded region (not visible in this
  plot) indicates these constraints, augmented by an estimate of the
  systematic errors in our fits; see \citet{PSutil2}.
These constraints provide no information about
  population synthesis.
}
\end{figure}

\subsubsection{Fitting results}
As in previous studies such as \citet{2005ApJ...620..385O},  here we have chosen to vary seven (7)
model parameters in the synthesis calculations.  These choices are strongly guided by our past
experience with double-compact-object population synthesis 
and represent the model parameters for which strong dependence has been confirmed. 
These seven parameters enter into every aspect of our population
synthesis model.  One parameter, a power law index $r$ in our
parameterization of the companion mass distribution, influences 
the initial binary parameter distributions (through the companion
masses).   Another parameter $w$, the massive stellar wind strength,
controls how rapidly the massive progenitors of compact objects lose
mass; this parameter strongly affects the final compact-object mass
distribution.  Three parameters $v_{1,2}$ and $s$ are used to
parameterize the supernovae kick distribution as a superposition of
two independent maxwellians.   These kicks provide critical opportunities to
push distant stars into tight orbits and also to disrupt
  potential double neutron star progenitors.    Finally, two parameters
$\alpha \lambda$ and $f_a$ govern  energy and mass
transfer during certain types of binary  interactions; see Section 2.2.4 of
\citet{2002ApJ...572..407B} for details.
To allow for an 
extremely broad range of possible models, we consider the specific
parameter ranges quoted in \S 2 of \citet{2005ApJ...620..385O} for all
parameters except kicks; for supernova kick parameters, we allow the
dispersion $v_{1,2}$ of either component of a bimodal Maxwellian to run from 0
to 1000 km/s.    

Since population synthesis calculations involve considerable
computational expense, in practice we estimate the merger rate we
expect for a given combination of population synthesis model
parameters via seven-dimensional fits to an archive of roughly 3000
detailed simulations, as presented and analyzed in
detail in \citet{PSutil2} (hereafter OKB).   However, these fits
introduce systematic errors, which have the potential to
significantly change the predicted set of constraint-satisfying
models.  For this reason, OKB also 
estimate the rms error associated with each seven-dimensional fit.
Finally, OKB demonstrate that, by broadening the constraint-satisfying
interval by the fit's rms error, 
the predicted set of constraint-satisfying models includes most models
which \emph{actually} satisfy the constraints.  For this reason, 
the dark shaded regions in Figures~\ref{fig:ConstraintsVersusDistribs:NSNS}, \ref{fig:ConstraintsVersusDistribs:WDNS}, and 
\ref{fig:ConstraintsVersusDistribs:SN}
 account for both
observational uncertainty in the Milky Way merger rate and for
systematic errors in the fits against which these observations will be compared.

\subsection{Prior distributions}

Lacking knowledge about which population synthesis model is correct,
we assume all population synthesis model parameters in our range are
\emph{equally likely}.  A  monte carlo over the seven-dimensional
parameter space allows us then to estimate the relative likelihood,
all things being equal, of various DCO merger rates (shown in 
Figures~\ref{fig:ConstraintsVersusDistribs:NSNS},\ref{fig:ConstraintsVersusDistribs:WDNS}) and supernova rates (Figure~\ref{fig:ConstraintsVersusDistribs:SN}).
Also shown in these figures are the observational constraints
described in Sections \ref{sec:dco:wdns}, \ref{sec:dco:nsns} and \ref{sec:data:other} (shown in shaded gray).
Finally, the dotted lines in Figure~\ref{fig:FinalRateDistribs} show the
a priori population synthesis predictions for BH-BH, BH-NS, and NS-NS
merger rates in the Milky Way.

\subsection{Applying and employing constraints}
\label{sec:constraints:DCO}

Physically consistent models must reproduce all predictions.
Supernovae rates, being ill-determined due to the large observational systematic
errors mentioned in Sec.~\ref{sec:data:other}, are easily reproduced
by almost all models at the $2\sigma$ level;
see Fig.~\ref{fig:ConstraintsVersusDistribs:SN}.  However, population
synthesis models do not always reproduce observed formation rates for
DCOs, as shown in Figs.~\ref{fig:ConstraintsVersusDistribs:WDNS} and
\ref{fig:ConstraintsVersusDistribs:NSNS}.  These four 95\% confidence
intervals implicitly define a $(0.95)^4\approx 81\%$
confidence interval in 
the seven-dimensional space, consisting of $9\%$ of the 
original parameter volume when systematic
errors in our fitting procedure are included.  In other words, we are
quite confident ($80\%$ chance) that the physically-appropriate
parameters entering into \emph{StarTrack} can be confined within a
small seven-dimensional volume, in principle significantly reducing
our model uncertainty.

\begin{figure}[h]
\begin{center}
\includegraphics[width=5.0in]{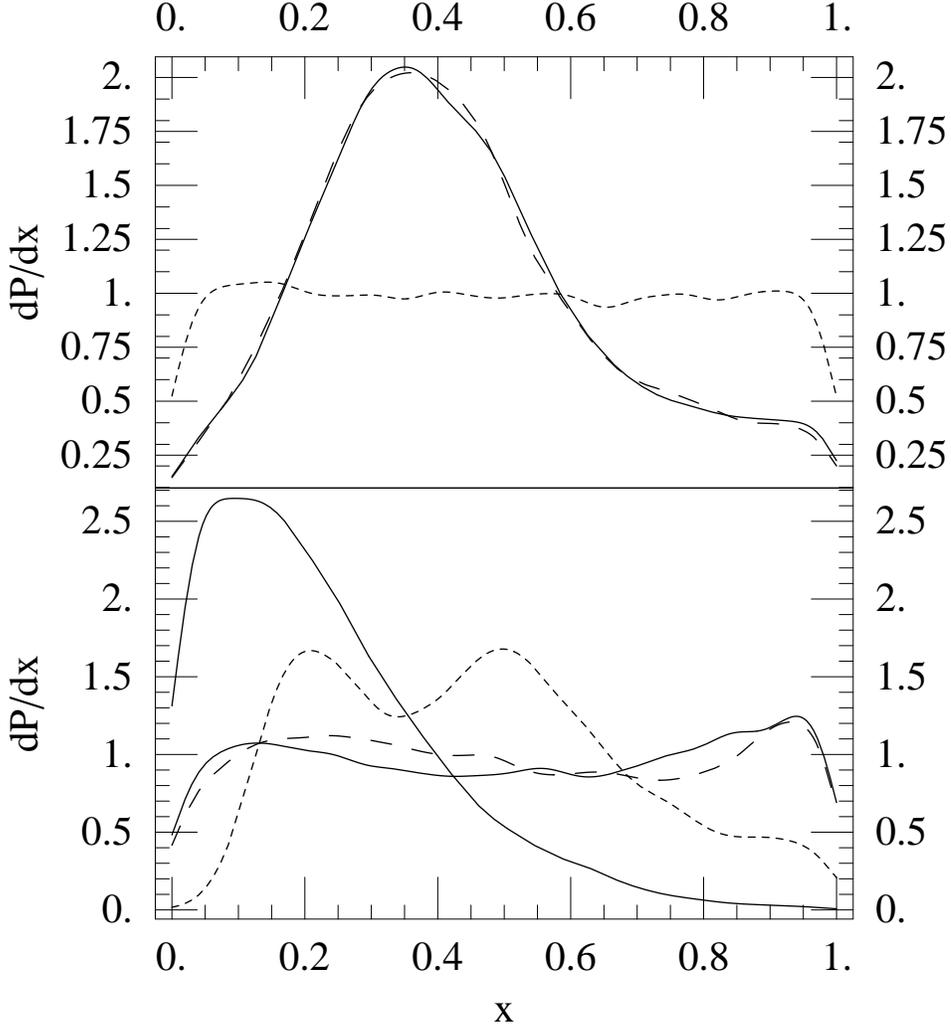}
\end{center}
\caption{\label{fig:params} Differential probability distributions
  $dP_k/dx$ defined so $P_k(x)$ is the fraction of all models
  consistent with all DCO observations and with 
  $x_k<X$.  The top panel shows the distributions for the 3
  kick-related parameters $x_3,x_4,x_5$ (dashed, solid, and dotted,
  respectively, corresponding to $v_1$, $v_2$, and $s$).  The bottom panel shows the
  distributions for  $x_1$ (the mass-ratio distribution parameter $r$, appearing as
  the solid nearly constant curve), and the three
  binary-evolution-related parameters $x_2,x_6,x_7$ (dashed, dotted,
  and solid, respectively, corresponding to $w$, $\alpha \lambda$, and
  $f_a$).
  The distribution of $f_a=x_7$ exhibits a strong peak somewhere
  between $0-0.1$; our choice of smoothing method causes all
  projected distributions to appear to drop to zero at the boundaries,
  as it involves averaging with empty cells.
}
\end{figure}

In Figure~\ref{fig:params} we show  one-dimensional projections onto
each coordinate axis of the constraint-satisfying volume -- in other
words, the distribution of values each \emph{individual} population
synthesis parameter can take.  For the
purposes of this and subsequent plots, we use the following seven
dimensionless parameters $x_k$ that run from 0 to 1: $x_1= a/3$; 
$x_2=w$; $x_3=v_1/(1000{\rm km/s})$, $x_4=v_2/(1000{\rm
  km/s})$; $x_5=s$; $x_6=\alpha\lambda$, and $x_7=f_a$.   
As seen in the top panel of this figure, the conditional distribution of supernovae kicks bears a surprising
resemblance to the observed pulsar kick distribution
\citep[see, e.g.,][and references
therein]{1999ApJ...520..696A, 2005MNRAS.360..974H, 2006ApJ...643..332F}, even though \emph{no
  information about pulsar motions have been included} among our
constraints and priors.  The population synthesis mass transfer
parameter $f_a$ is also well-constrained, with a strong maximum near
$x=0$,
implying that mass-loss fractions of about 90\% or higher are favored
in non-conservative, but stable, mass transfer episodes. Also, common
envelope efficiencies $\alpha\lambda$ of about 0.2-0.5 appear to be
favored by the constraints. The rest of the one-dimensional parameter
distributions are nearly constant and thus uninformative.    
Though small, the
constraint-satisfying volume extends throughout the seven-dimensional
parameter space.

\begin{figure}
\begin{center}
\includegraphics[width=4.in]{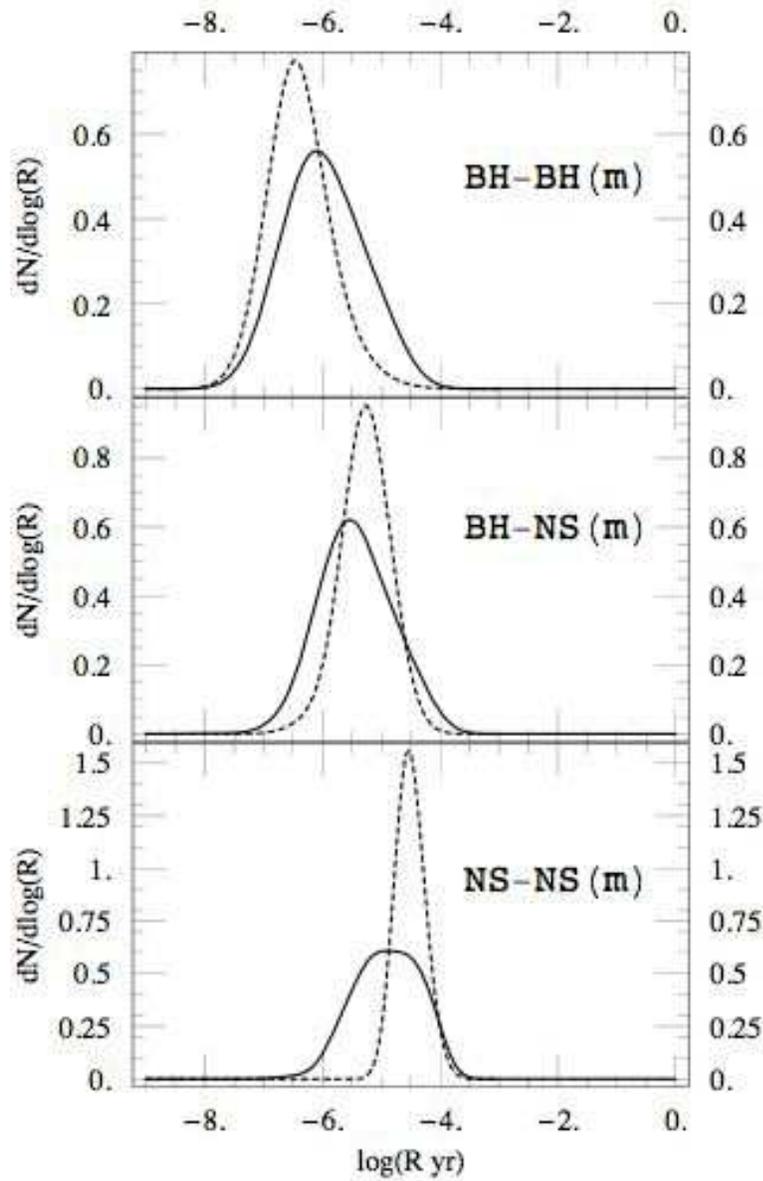}
\end{center}
\caption{\label{fig:FinalRateDistribs} A plot of the  probability distributions for the BH-BH
(top), BH-NS (center), and NS-NS (bottom) merger rates per
Milky Way-equivalent galaxy, allowing for systematic errors
in the BH-BH, BH-NS, and NS-NS fits.   The solid
 curves result from smoothing a histogram of results from a random
 sample of population synthesis calculations with  systematic errors included.
The dashed  curve  shows
 the same results, assuming model parameters are restricted to those
 which satisfy all four DCO constraints (both WD-NS
and both NS-NS constraints).   Compare with Figure 5 of
\citet{2005ApJ...633.1076O}.  
}
\end{figure}

\begin{figure}
\begin{center}
\includegraphics[width=5.0in]{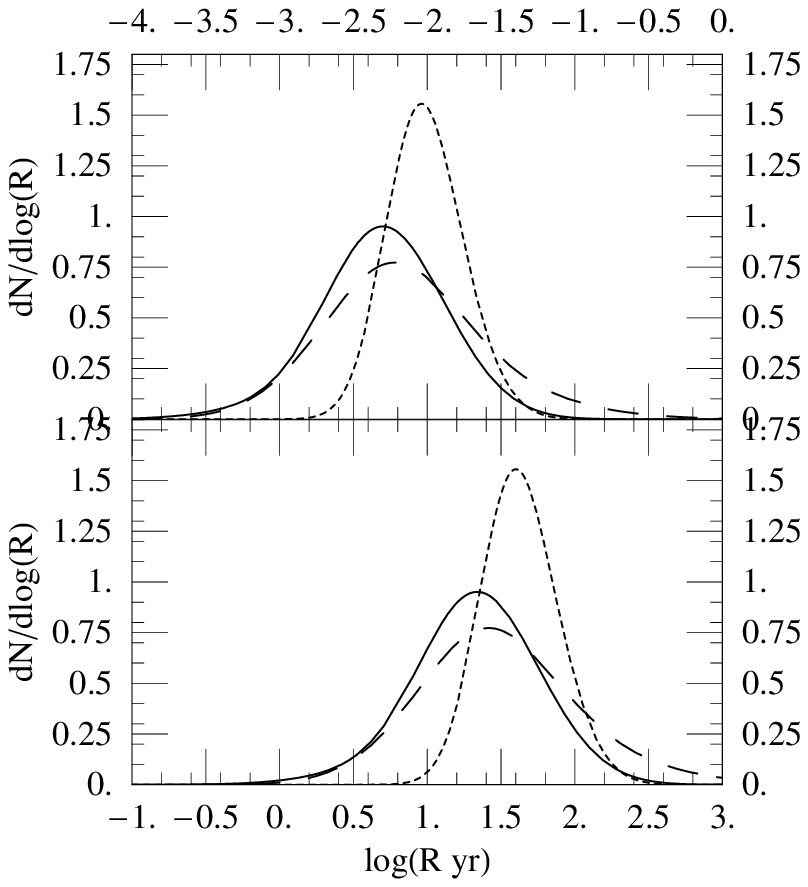}
\end{center}
\caption{\label{fig:results:LIGO1} Range of expected detection rates for
  initial (top) and advanced (bottom) LIGO for BH-BH (dashed), BH-NS (solid), and NS-NS (dotted)
  binaries, based on models that satisfy observational constraints and
  under the assumptions outlined in \citet{2005ApJ...633.1076O} (e.g., using
  a fixed chirp-mass spectrum and LIGO range for detection, as well as assuming Milky-Way like
  galaxies fill the universe with density $0.01 {\rm Mpc}^{-3}$). 
  These estimates incorporate systematic errors due to the fitting
  procedure.  
}
\end{figure}

Evaluating fits for other DCO merger
rates  on the constraint-satisfying
model parameters provides  revised estimates for various phenomena
of interest, such as for the BH-BH, BH-NS, and NS-NS merger rates.
Figure~\ref{fig:FinalRateDistribs} shows smoothed
histograms of the BH-BH, BH-NS, and NS-NS  merger rates, both before and
after observational constraints were applied.   Additionally, a
smoothed curve shows the results modulated by the effect of systematic
error.
These results should
be contrasted with the distinctly different post-constraint results
shown in \citet{2005ApJ...633.1076O} (cf. their Figure 5 and C7).  Notably, even
though \emph{two fewer} constraints were imposed, they find a
significantly smaller ($2\%$ versus $9\%$) constraint-satisfying
volume. 
 In large part, these
differences can be ascribed to including systematic
errors: in \citet{2005ApJ...633.1076O}, preliminary data for
the sparse and poor-quality NSNS(vw) sample was fit and applied without any
account for fit-induced errors.   However, our calculation also
differs at several fundamental levels from the original approach :
(i) the space of  models studied is larger, covering more area in
$v_1,v_2$ \citep[see,e.g.,][]{PSutil2}; 
(ii) the observed sample of merging NS-NS binaries is compared with
the \emph{total} merging NS-NS formation rate predicted from
population synthesis, rather than with subset of merging NS-NS which
undergo mass transfer; and critically (iii) the constraints are more numerous.

Finally,  in Figure \ref{fig:results:LIGO1} we  estimate the range of initial
and advanced LIGO detection rates implied by these constrained
results.  This estimate combines our constrained population synthesis
results; our estimates for systematic fitting error; the approximate LIGO range as
a function of chirp mass $M_c$:
$
d = d_o \left({\cal M}_c/ 1.2 M_\odot \right)^{5/6}
$ where $d_o=15\sqrt{3/2}$, $300\, {\rm Mpc}$ for initial and advanced
LIGO, respectively; estimates for the average volume detected for a
given species, as represented by 
$\left<M_c^{15/6}\right>=111,5.8,2 M_\odot^{15/6}$ for BH-BH, BH-NS,
and NS-NS, respectively \citep[see][]{2005ApJ...633.1076O};
and a homogeneous,
Euclidean universe  populated with Milky Way-equivalent galaxies with
density $0.01 {\rm Mpc}^{-3}$, each forming stars at rate $3.5
M_\odot{\rm yr}^{-1}$.  Updated rates that address a number of these
simplifications will be examined in two forthcoming papers,
\cite{PSgrbs-popsyn} and \cite{PSellipticals}.

\subsection{Conclusions}

\citet{2005ApJ...633.1076O}  performed a
proof-of-concept calculation to compare population synthesis
simulations with observations, interpret the resulting constrained
volume, and apply the constraint-satisfying parameters to revise a
priori predictions of astrophysically critical rate results.
Here we demonstrate that even when 
systematic  model fitting errors are aggressively overcompensated  for, 
observational constraints still provide information about population
synthesis parameters.   These results include both surprisingly good
agreement with pulsar kick distributions and strong constraints on 
at least one
parameter involved in binary evolution.   Since in almost
all cases our systematic errors appear much smaller than those from
observations, we are confident that now \emph{observational} limitations,
rather than computational ones, primarily limit  our ability to constrain
population synthesis results.   In particular, we expect stringent
tests of  population synthesis models are possible, beginning by
imposing more  observational constraints than varied model parameters
(seven in our case).

Our approach also strongly suggests  that population synthesis
estimates for DCO merger rates vary in a very complex manner, as
indicated by the large $\chi^2$  produced by even the best  low-order
polynomial fits to often often very high-quality data \citep[see, e.g.][]{PSutil2}.  We strongly
suspect that a more physically motivated class of basis functions
could achieve a much better fit; we intend to explore this
possibility, along with the reliability of purely nonparametric fits,
in a future study.

Our analysis remains predicated upon a correct identification of all parameters and input physics critical for the formation of double compact objects. Admittedly, our understanding of binary evolution continues to evolve; however multiple population studies from different groups over the years lead to very similar conclusions about the basic input physics that primarily affects the formation rates of double compact objects~\citep{2002ApJ...572..407B, 2002MNRAS.329..897H, 1996AA...309..179P, 1998ApJ...496..333F}. In this paper we use a relatively new version of the \emph{StarTrack} code that is described in great detail in \citet{2005astro.ph.11811B}. A forthcoming paper (Belczynski, Kalogera, \& Bulik, in preparation) will discuss how these updates affect the evolutionary history and formation rates in  significantly greater detail.
  
Our analysis also remains predicated upon a correct and complete
interpretation of 
the observational sample and associated systematic and model
uncertainties. In some respects, however, our models for observations 
and their biases may be incomplete or
not representative. 
We effectively assume pulsar recycling is required to detect binary
pulsars in our implicit hypothesis 
that only one pulsar in NS-NS systems, the one observed,
was ever likely to have been detected.    And finally we use a
canonical value for pulsar beaming correction factor $f_a$ for several
systems with known spins but unknown beaming geometry.  (We hope to
address these questions in a forthcoming paper.)

Last, our analysis of DCO merger rates remains predicated on the assumption that the
Milky Way formed stars at a relatively uniform rate for the last
$10$~Gyr.  However, as will be demonstrated at greater length in two
forthcoming papers \citep{PSellipticals, PSgrbs-popsyn}, if the star
formation rate increased significantly
when the Milky Way was young, then since most  binaries would then form
early in the lifetime of the Milky Way, the young-galaxy contribution
could conceivably overwhelm the present-day contribution.  In other
words, the present-day Milky Way merger rate for BH-BH binaries, for
example, could be completely determined by the star formation rate
nearly $10$~Gyr ago.

And to be comprehensive regarding the factors neglected here, our
population synthesis parameter study has not varied over \emph{all}
plausible models.  We have neither considered purely polar supernova kicks
\citep[see][]{2005MNRAS.364.1397J} nor changed the maximum NS mass
from $2 M_\odot$ nor even allowed for a heterogeneous birth population of single stars 
and binaries (a binary fraction of 100\% is assumed in all models).  And of course we continue to use a single fixed star
formation rate for the Milky Way, rather than treat it as an
independent uncertain but constrained parameter.

However, all these limitations can be remedied by closer study.  
Thus, we  expect that either (i) all DCO merger
rates will become determined to within $O(50\%)$ or better (based on
the comparative accuracy to which we know NS-NS merger rates), or (ii)
experimental data will conflict with the prevailing notion of binary
population synthesis, revealing flaws and limitations in classical
parametric models for binary stellar evolution.

\section{Open Issues} 
\label{end}

It has now been more than 30 years since the discovery of the first binary pulsar that happened to also be the first binary with two neutron stars. This discovery has essentially led to the development of a whole sub-field in compact object astrophysics and progress in this area has also greatly contributed to the progress in the development of gravitational-wave observatories over these years.

As we anticipate the direct detection of gravitational waves from binary inspirals and the possible, associated discovery of black hole binaries, we also take advantage of pulsar timing observations to the fullest and continually try to improve our understanding of the origin of tight double compact objects. It is clear that a lot of progress has been particularly made in the last few years with the discovery of the highly relativistic, first double pulsar and additional double neutron stars, and the tremendous progress in proper motion and pulse profile evolution measurements. More interestingly, our standard way of thinking about double compact object formation is called to question with a number of intriguing suggestions and counter-suggestions put forward: do neutron stars form through two different physical mechanisms depending on the mass of their immediate progenitor at the time of core collapse \citep{2005MNRAS.361.1243P, 2005PhRvL..94e1102P, 2006PhRvD..74d3003W, 2006MNRAS.tmpL.105S} do some or the majority of neutron stars receive minimal supernova kicks with magnitudes significantly smaller than 50\,km\,s$^{-1}$ \citep{2004inun.conf..185V, 2005ApJ...632.1054C, 2005astro.ph..8626I, 2005MNRAS.363L..71D}. Furthermore, some of the older questions still persist: what is the basic physical mechanism for core collapse that also drives supernovae? Do most black holes form through fallback onto proto-neutron-stars and supernovae or through direct collapse? What is the mass function of black holes and how does it depend on metallicity? Why are neutron star masses so close to one another in double neutron stars? What really determines the birth spins and spin evolution of neutron stars and black holes? 

Starting with the very basic question of the core collapse physical mechanism Hans Bethe was intrigued by many of these questions and naturally his interest evolved towards binary compact objects, their evolutionary history, and their potential as gravitational wave sources. Given the progress we have experienced since the discovery of the first DNS binary, it is reasonable to expect that with the combination of developments in pulsar searches and observations, of the rise of gravitational-wave detections and astrophysics, and of the progress in theoretical physical understanding and computational astrophysics we will be able to answer some of the basic remaining questions raised by the existence of binary compact objects in nature.

\noindent
{\bf Acknowledgements}
 The work summarized here has been supported by a wide range of research funding sources. VK is grateful for support from NSF Gravitational Physics grants No. PHYS-0121416 and PHY-0353111, a David and Lucile Packard Foundation Fellowship in Science and Engineering grant, and a Cottrell Scholar Award from the Research Corporation. 
 K.B. acknowledges the support of KBN Grant No. 1P03D02228.


\newpage 

\begin{thebibliography}{}


\bibitem[Abbott et al.(2005)]{2005PhRvD..72h2002A}
Abbott, B., et al.\ 2005, Phys.\ Rev.\ D, 72, 082002

\bibitem[Abramovici et al.(1992)]{1992Sci...256..325A}
Abramovici, A., et al.\ 1992, Science, 256, 325

\bibitem[Arzoumanian et al.(1999)]{1999ApJ...520..696A} 
Arzoumanian, Z., Cordes, J.~M., \& Wasserman, I.\ 1999, ApJ, 520, 696 

\bibitem[Bailes et al.(2003)]{2003ApJ...595L..49B}
Bailes, M., Ord, S.~M., Knight, H.~S., \& Hotan, A.~W. 2003, ApJ, 595, L49

\bibitem[Belczy{\'n}ski \& Kalogera(2001)]{2001ApJ...550L.183B}  
Belczy{\'n}ski, K., \& Kalogera, V.\ 2001, ApJ Letters, 550, L183   

\bibitem[Belczynski et al.(2002a)]{2002ApJ...571L.147B} 
Belczynski, K., Bulik, T., \& Kalogera, V.\ 2002, ApJ Letters, 571, L147

\bibitem[Belczynski et al.(2002b)]{2002ApJ...572..407B} 
Belczynski, K., Kalogera, V., \& Bulik, T.\ 2002, ApJ, 572, 407

\bibitem[Belczynski et al.(2006a)]{2006ApJ...648.1110B}
Belczynski, K., Perna, R., Bulik, T., Kalogera, V., Ivanova, N., \& Lamb, D.~Q.\ 2006, 
ApJ, 648, 1110

\bibitem[Belczynski et al.(2006b)]{2005astro.ph.11811B} 
Belczynski, K., Kalogera, V., Rasio, F.~A., Taam, R.~E., Zezas, A., Bulik, T.,
  Maccarone, T.~J., \& Ivanova, N.\ 2006, ArXiv Astrophysics e-prints,
  arXiv:astro-ph/0511811 

\bibitem[Bethe \& Brown(1998)]{1998ApJ...506..780B}
Bethe, H.~A., \& Brown, G.~E.\ 1998, ApJ, 506, 780

\bibitem[Bethe \& Brown(1999)]{1999ApJ...517..318B}
Bethe, H.~A., \& Brown, G.~E.\ 1999, ApJ, 517, 318

\bibitem[Bhattacharya \& van den Heuvel(1991)]{1991PhR...203....1B}
  Bhattacharya, D., \& van den Heuvel, E.~P.~J.\ 1991, Physics
  Reports, 203, 1 

\bibitem[Brandt \& Podsiadlowski(1995)]{1995MNRAS.274..461B}
Brandt, N., \& Podsiadlowski, P.\ 1995, Monthly Notices of the Royal
  Astronomical Society, 274, 461

\bibitem[Brown(1995)]{1995ApJ...440..270B} 
Brown, G.~E.\ 1995, ApJ, 440, 270 

\bibitem[Burgay et al.(2003)]{2003Natur.426..531B}
Burgay, M., et al.\ 2003, Nature, 426, 531 
  
\bibitem[Camilo et al.(2002)]{2002ApJ...567L..71C}
Camilo, F., Manchester, R.~N., Gaensler, B.~M., Lorimer, D.~R., \& Sarkissian, J.\ 2002, ApJ Letters, 567, L71
  
\bibitem[Camilo(2003)]{2003ASPC..302..145C} 
Camilo, F.\ 2003, in {\it Radio Pulsars} (ASP Conf. Ser.), eds.\ M.\ Bailes,  D.J.\ Nice, \& S.E.\ Thorsett (San Francisco), 302, 145

\bibitem[Cappellaro et al.(1999)]{1999A&A...351..459C} 
Cappellaro, E., Evans, R., \& Turatto, M.\ 1999, A\&A, 351, 459

\bibitem[Cappellaro et al.(2005)]{2005A&A...430...83C}
Cappellaro, E., Riello, M., Altavilla, G., Botticella, M.~T.,
  Benetti, S., Clocchiatti, A., Danziger, J.~I., Mazzali, P.,
  Pastorello, A., Patat, F., Salvo, M., Turatto, M., \& Valenti, S.\
  2005, A\&A, 430, 83

\bibitem[Carlberg \& Innanen(1987)]{1987AJ.....94..666C} 
Carlberg, R.~G., \& Innanen, K.~A.\ 1987, Astronomical Journal, 94, 666 
  
\bibitem[Champion et al.(2004)]{2004MNRAS.350L..61C}
Champion, D.~J., Lorimer, D.~R., McLaughlin, M.~A., Cordes, J.~M., Arzoumanian, Z., Weisberg, J.~M. \& Taylor, J.~H.\ 2004, MNRAS, 350, L61

\bibitem[Chatterjee et al.(2005)]{2005ApJ...630L..61C} 
Chatterjee, S., et al.\ 2005, ApJ Letters, 630, L61

\bibitem[Chaurasia and Bailes(2005)]{2005ApJ...632.1054C}
Chaurasia, H.~K., \& Bailes, M.\ 2005, ApJ, 632, 1054

\bibitem[Chen \& Ruderman(1993)]{1993ApJ...402..264C}
Chen, K. \& Ruderman, M.\ 1993, ApJ, 402, 264

\bibitem[Chen et al.(2001)]{2001ApJ...553..184C} 
Chen, B., et al.\ 2001, ApJ, 553, 184
  
\bibitem[Coles et al.(2005)]{2005ApJ...623..392C} 
Coles, W.~A., McLaughlin, M.~A., Rickett, B.~J., Lyne, A.~G., \& Bhat, N.~D.~R.\
  2005, ApJ, 623, 392
  
\bibitem[Cordes et al.(1993)]{1993Natur.362..133C} 
Cordes, J.~M., Romani, R.~W., \& Lundgren, S.~C.\ 1993, Nature, 362, 133

\bibitem[Cordes \& Chernoff(1997)]{1997ApJ...482..971C} 
Cordes J.M. \& Chernoff, D.F.\ 1997, ApJ, 482, 971 

\bibitem[Cox(2000)]{2000asqu.book.....C}
Cox, A.~N.\ 2000, Allen's Astrophysical Quantities, 4th edn.

\bibitem[Demorest et. al(2004)]{2004ApJ...615L.137D} 
Demorest, P., Ramachandran, R., Backer, D.C., Ransom,
  S.M., Kaspi, V., Arons, J., \& Spitkovsky, A.\ 2004, ApJ Letters, 615, L137

\bibitem[Dewi et al.(2002)]{2002MNRAS.331.1027D} 
Dewi, J.~D.~M., Pols, O.~R., Savonije, G.~J., \& van den Heuvel, E.~P.~J.\ 2002, Monthly
  Notices of the Royal Astronomical Society, 331, 1027

\bibitem[Dewi \& Pols(2003)]{2003MNRAS.344..629D} 
Dewi, J.~D.~M., \& Pols, O.~R.\ 2003, Monthly Notices of the Royal Astronomical
  Society, 344, 629

\bibitem[Dewi \& van den Heuvel(2004)]{2004MNRAS.349..169D} 
Dewi, J.~D.~M., \& van den Heuvel, E.~P.~J.\ 2004, Monthly Notices of the Royal
  Astronomical Society, 349, 169

\bibitem[Dewi et al.(2005)]{2005MNRAS.363L..71D}
Dewi, J.~D.~M., Podsiadlowski, Ph., Pols, O.~R.\ 2005, Monthly 
Notices of the Royal Astronomical Society, 363, L71

\bibitem[Dewi et al.(2006)]{Sena}
Dewi, J.D.M., Podsiadlowski, Ph., Sena, A.\ 2006, Monthly 
Notices of the Royal Astronomical Society, 368, 1742

\bibitem[Diehl et al.(2006)]{2006Natur.439...45D}
Diehl, R., Halloin, H., Kretschmer, K., Lichti, G.~G.,
  Sch{\"o}nfelder, V., Strong, A.~W., von Kienlin, A., Wang, W.,
  Jean, P., Kn{\"o}dlseder, J., Roques, J.-P., Weidenspointner, G.,
  Schanne, S., Hartmann, D.~H., Winkler, C., \& Wunderer, C.\ 2006,
  Nature, 439, 45

\bibitem[Edwards \& Bailes(2001)]{2001ApJ...547L..37E}
Edwards, R.~T. \& Bailes, M.\ 2001, ApJ Letters, 547, L37

\bibitem[Faucher-Giguere \& Kaspi(2006)]{2006ApJ...643..332F}
  Faucher-Giguere, C.~-A., \& Kaspi, V.~M.\ 2006, ApJ, 643, 332
  
\bibitem[Faulkner et al.(2005)]{2005ApJ...618L.119F}
Faulkner, A.~J., Kramer, M., Lyne, A.~G., Manchester, R.~N.,
McLaughlin, M.~A., Stairs, I.~H., Hobbs, G., Possenti, A., Lorimer,
 D.~R., D'Amico, N., Camilo, F., \& Burgay, M.\ 2005, ApJ Letters, 618, L119

\bibitem[Fryer \& Kalogera(1997)]{1997ApJ...489..244F} 
Fryer, C., \& Kalogera, V.\ 1997, ApJ, 489, 244

\bibitem[Fryer et al.(1998)]{1998ApJ...496..333F}
Fryer, C., Burrows, A., \& Benz, W.\ 1998, ApJ, 496, 333 

\bibitem[Habets(1986)]{1986A&A...167...61H} 
Habets, G.~M.~H.~J.\ 1986, Astronomy \& Astrophysics, 167, 61

\bibitem[Hills(1983)]{1983ApJ...267..322H} 
Hills, J.~G.\ 1983, ApJ, 267, 322

\bibitem[Hobbs et al.(2005)]{2005MNRAS.360..974H} 
Hobbs, G., Lorimer, D.~R., Lyne, A.~G., \& Kramer, M.\ 2005, Monthly Notices of the
  Royal Astronomical Society, 360, 974
  
\bibitem[Hulse \& Taylor(1975)]{1975ApJ...195L..51H}
Hulse R. \& Taylor, J.~H.\ 1975, ApJ, 195, L51

\bibitem[Hurley et al.(2002)]{2002MNRAS.329..897H}
Hurley, J.~R., Tout, C.~A., \& Pols, O.~R.\ 2002, MNRAS, 329, 897

\bibitem[Ihm et al.(2006)]{2005astro.ph..8626I}
Ihm, C.~M., Kalogera, V., \& Belczynski, K.\ 2006, ApJ, in press [astro-ph/0508626]

\bibitem[Ivanova et al.(2003)]{2003ApJ...592..475I} 
Ivanova, N., Belczynski, K., Kalogera, V., Rasio, F.~A., \& Taam, R.~E.\ 2003,
  ApJ, 592, 475

\bibitem[Jenet \& Ransom(2004)]{20042004Natur.428..919J}
Jenet, F.A. \& Ransom, S.M.\ 2004, Nature, 428, 919

\bibitem[Johnston et al.(2005)]{2005MNRAS.364.1397J}
Johnston, S., Hobbs, G., Vigeland, S., Kramer, M., Weisberg, J.~M.,
  \& Lyne, A.~G.\ 2005, MNRAS, 364, 1397

\bibitem[Junker \& Sch{\"a}fer(1992)]{1992MNRAS.254..146J} 
Junker, W., \& Sch{\"a}fer, G.\ 1992, Monthly Notices of the Royal Astronomical
  Society, 254, 146 

\bibitem[Kalogera(1996)]{1996ApJ...471..352K} 
Kalogera, V.\ 1996, ApJ, 471, 352 

\bibitem[Kalogera \& Lorimer(2000)]{2000ApJ...530..890K} 
Kalogera, V., \& Lorimer, D.~R.\ 2000, ApJ, 530, 890 
  
\bibitem[Kalogera(2000)]{2000ApJ...541..319K} 
Kalogera, V.\ 2000, ApJ, 541, 319  

\bibitem[Kalogera et al.(2001)]{2001ApJ...556..340K} Kalogera, V.,
  Narayan, R., Spergel, D.~N., \& Taylor, J.~H.\ 2001, ApJ, 556, 340  

\bibitem[Kalogera et al.(2004a)]{2004ApJ...601L.179K} 
Kalogera, V., et al.\ 2004, ApJ Letters, 601, L179
  
\bibitem[Kalogera et al.(2004b)]{2004ApJ...614L.137K} Kalogera, V., et
  al.\  2004, ApJ Letters, 614, L137

\bibitem[Kalogera et al.(2005)]{2005ASPC..328..261K}
Kalogera, V., Kim, C., Lorimer, D.~R., Ihm, M., \& Belczynski, K.\
 2005, in ASP Conf. Ser. 328: Binary Radio Pulsars, ed. I.~Stairs, 261
  
\bibitem[Kaspi et al.(2000)]{2000ApJ...543..321K}
Kaspi, V.~M., Lyne, A.~G., Manchester, R.~N., Crawford, F., Camilo,
  F., Bell, J.~F., D'Amico, N., Stairs, I.~H., McKay, N.~P.~F.,
  Morris, D.~J., \& Possenti, A.\ 2000, ApJ, 543, 321

\bibitem[Kim et al.(2003)]{2003ApJ...584..985K}
Kim, C., Kalogera, V., \& Lorimer, D.~R.\ 2003, ApJ, 584, 985 (KKL)

\bibitem[Kim et al.(2004)]{2004ApJ...616.1109K} 
Kim, C., Kalogera, V., Lorimer, D.~R., \& White, T.\ 2004, ApJ, 616, 1109 

\bibitem[Kim et al.(2006)]{2006.astro-ph..0608280}
Kim, C., Kalogera, V., \& Lorimer, D.~R.\ 2006, in A life with stars [astro/ph/0608280]

\bibitem[Kramer(1998)]{1998ApJ...509..856K}
Kramer, M.\ 1998, ApJ, 509, 856

\bibitem[Kramer et al.(2003)]{2003ApJ...593L..31K} 
Kramer, M., Lyne, A.~G., Hobbs, G., L{\" o}hmer, O., Carr, P., Jordan, C., \&
  Wolszczan, A.\ 2003, ApJ Letters, 593, L31 

\bibitem[Kramer et al.(2005)]{2005astro.ph..3386K} 
Kramer, M., et al.\ 2005, ArXiv Astrophysics e-prints, arXiv:astro-ph/0503386

\bibitem[Kramer et al.(2006)]{2006Sci...314...97K}
Kramer, M., Stairs, I.~H., Manchester, R.~N., McLaughlin, M.~A., Lyne, A.~G., Ferdman, R.~D., Burgay, M., Lorimer, D.~R., Possenti, A., D'Amico, N., Sarkissian, J.~M., Hobbs, G.~B., Reynolds, J.~E., Freire, P.~C.~C., \& Camilo, F.\ 2006, Science, 314, 97

\bibitem[Kuijken \& Gilmore(1989)]{1989MNRAS.239..571K} 
Kuijken, K., \& Gilmore, G.\ 1989, Monthly Notices of the Royal Astronomical
  Society, 239, 571
  
\bibitem[Lorimer et al.(2005)]{2005ASPC..328..113L} 
Lorimer, D.~R., et al.\ 2005, in {\it Binary Radio Pulsars} (ASP Conf. Ser.), eds.\ F.~A.\ Rasio \& I.~H.\ Stairs (San Francisco), 328, 113 

\bibitem[Lorimer et al.(2006)]{2006ApJ...640..428L}
Lorimer, D.~R., et al.\ 2006, ApJ, 640, 428

\bibitem[Lyne et al.(2004)]{2004Sci...303.1153L} 
Lyne, A.~G., Burgay, M., Kramer, M., Possenti, A., Manchester, R.~N.,
  Camilo, F., McLaughlin, M.~A., Lorimer, D.~R., D'Amico, N., Joshi,
  B.~C., Reynolds, J., \& Freire, P.~C.~C.\ 2004, Science, 303, 1153
  
\bibitem[Lyutikov(2004)]{2004MNRAS.353.1095L} 
Lyutikov, M.\ 2004, Monthly Notices of the Royal Astronomical Society, 353, 1095

\bibitem[Lyutikov \& Thompson(2005)]{2005ApJ...634.1223L} 
Lyutikov, M., \& Thompson, C.\ 2005, ApJ, 634, 1223 
 
\bibitem[Manchester et al.(2005)]{2005ApJ...621L..49M} 
Manchester, R.~N., et al.\ 2005, ApJ Letters, 621, L49 

\bibitem[Narayan et al.(1991)]{1991ApJ...379L..17N}
Narayan, R., Piran, T., \& Shemi, A.\ 1991, ApJ Letters, 379, L17

\bibitem[Nomoto(1984)]{1984ApJ...277..791N} Nomoto, K.\ 1984, ApJ,
  277, 791
   
\bibitem[Nomoto(1987)]{1987ApJ...322..206N} Nomoto, K.\ 1987, ApJ,
  322, 206 

\bibitem[O'Shaughnessy et al.(2005a)]{2005ApJ...620..385O}
O'Shaughnessy, R., Kalogera, V., \& Belczynski, K.\ 2005,
 ApJ, 620, 385
 
\bibitem[O'Shaughnessy et al.(2005b)]{2005ApJ...633.1076O}
O'Shaughnessy, R., Kim, C., Fragos, T., Kalogera, V., \& Belczynski,
  K.\ 2005, ApJ, 633, 1076
  
\bibitem[O'Shaughnessy et al.(2006a)]{PSutil2}
O'Shaughnessy, R., Kalogera, V., \& Belczynski\ 2006, submitted to ApJ [astro/ph/0609465]

\bibitem[O'Shaughnessy et al.(2006b)]{PSellipticals}
O'Shaughnessy, R., Kalogera, V., \& Belcynski, K.\ 2006, in
  preparation
  
\bibitem[O'Shaughnessy et al.(2006c)]{PSgrbs-popsyn}
O'Shaughnessy, R., Belczynski, C., Kalogera,
  V., \& Lamb, D.\ 2006, in preparation

\bibitem[Paczynski(1990)]{1990ApJ...348..485P} 
Paczynski, B.\ 1990, ApJ, 348, 485

\bibitem[Peters \& Mathews(1963)]{pm63} 
Peters P.~C. \& Mathews, J.\ 1963, Phys.\ Rev.\ D, 131, 435

\bibitem[Pfahl et al.(2002)]{2002ApJ...574..364P} 
Pfahl, E., Rappaport, S., Podsiadlowski, P., \& Spruit, H.\ 2002, ApJ, 574, 364 

\bibitem[Phinney(1991)]{1991ApJ...380L..17P} 
Phinney, E.~S.\ 1991, ApJ, 380, L17 

\bibitem[Piran \& Shaviv(2005a)]{2005PhRvL..94e1102P} Piran, T., \& 
  Shaviv, N.~J.\ 2005, Physical Review Letters, 94, 051102

\bibitem[Piran \& Shaviv(2005b)]{2005astro.ph.10584P} Piran, T., \&
  Shaviv, N.~J.\ 2005, ArXiv Astrophysics e-prints,
  arXiv:astro-ph/0510584

\bibitem[Piran \& Shaviv(2006)]{2006astro.ph..3649P} Piran, T., \& Shaviv, 
N.~J.\ 2006, ArXiv Astrophysics e-prints, arXiv:astro-ph/0603649

\bibitem[Pinsonneault \& Stanek(2006)]{twin}
Pinsonneault, M.H.\ \& Stanek, K.Z.\ 2006, ApJ, 639, L67

\bibitem[Podsiadlowski et al.(2005)]{2005MNRAS.361.1243P}
  Podsiadlowski, P., Dewi, J.~D.~M., Lesaffre, P., Miller, J.~C.,
  Newton, W.~G., \& Stone, J.~R.\ 2005, Monthly Notices of the Royal
  Astronomical Society, 361, 1243

\bibitem[Portegies Zwart \& Verbunt(1996)]{1996AA...309..179P}
Portegies Zwart, S.~F. \& Verbunt, F. 1996, A\&A, 309, 179

\bibitem[Portegies Zwart \& Yungelson(1998)]{1998AA...332..173P}
Portegies Zwart, S.~F. \& Yungelson, L.~R.\ 1998, A\&A , 332, 173

\bibitem[Ransom et al.(2004)]{2004ApJ...609L..71R} 
Ransom, S.~M., Kaspi, V.~M., Ramachandran, R., Demorest, P., Backer, D.~C., Pfahl,
  E.~D., Ghigo, F.~D., \& Kaplan, D.~L.\ 2004, ApJ Letters, 609, L71 

\bibitem[Stairs et al.(2004)]{2004PhRvL..93n1101S} 
Stairs, I.~H., Thorsett, S.~E., \& Arzoumanian, Z.\ 2004, Phys.\ Rev.\ Lett., 93, 141101 

\bibitem[Stairs et al.(2006)]{2006MNRAS.tmpL.105S}
Stairs, I.~H., Thorsett, S.~E., Dewey, R.~J., Kramer, M., \& McPhee, C.~A.\ 
2006, MNRAS, in press [astro-ph/0609416]

\bibitem[Stokes et al.(1985)]{1985ApJ...294L..21S} 
Stokes, G.~H., Taylor, J.~H., \& Dewey, R.~J.\ 1985, ApJ, 294, L21

\bibitem[Tauris \& van den Heuvel(2006)]{2003astro.ph..3456T} 
Tauris, T.~M., \& van den Heuvel, E.\ 2006, ArXiv Astrophysics e-prints,
  arXiv:astro-ph/0303456   

\bibitem[Thorsett et al.(2005)]{2005ApJ...619.1036T}
Thorsett, S.~E., Dewey, R.~J., \& Stairs, I.~H.\ 2005, ApJ, 619, 1036

\bibitem[van den Bergh \& Tammann(1991)]{1991ARAA..29..363V}
van den Bergh, S. \& Tammann, G.~A.\ 1991, ARA\&A, 29, 363
  
\bibitem[van den Heuvel(2004)]{2004inun.conf..185V}
van den Heuvel,  E.~P.~J.\ 2004, Proceedings of 5th INTEGRAL Workshop on the INTEGRAL Universe, 185  

\bibitem[van Kerkwijk \& Kulkarni(1999)]{1999ApJ...516L..25V}
van Kerkwijk, M.~H. \& Kulkarni, S.~R.\ 1999, ApJ, 516, L25

\bibitem[Weisberg \& Taylor(2002)]{2002ApJ...576..942W} 
Weisberg, J.\ \& Taylor, J.~H.\ 2002, ApJ, 576, 942 

\bibitem[Wex et al.(2000)]{2000ApJ...528..401W}
Wex, N., Kalogera, V., \& Kramer, M.\ 2000, ApJ, 528, 401

\bibitem[Willems \& Kalogera(2004)]{2004ApJ...603L.101W} 
Willems, B., \& Kalogera, V.\ 2004, ApJ Letters, 603, L101

\bibitem[Willems et al.(2004)]{2004ApJ...616..414W} Willems, B.,
  Kalogera, V., \& Henninger, M.\ 2004, ApJ, 616, 414
  
\bibitem[Willems et al.(2005)]{2005ApJ...625..324W} Willems, B.,
  Henninger, M., Levin, T., Ivanova, N., Kalogera, V., McGhee, K.,
  Timmes, F.~X., \& Fryer, C.~L.\ 2005, ApJ, 625, 324

\bibitem[Willems et al.(2006)]{2006PhRvD..74d3003W} 
Willems, B., Kaplan, J., Fragos, T., \& Kalogera, V.\ 2006, Phys.\ Rev.\ D, 74, 043003

\bibitem[Wolszczan(1991)]{1991Natur.350..688W}
Wolszczan, A.\ 1991, Nature, 350, 688

\bibitem[Zou et al.(2005)]{2005MNRAS.362.1189Z} Zou, W.~Z., Hobbs, G.,
  Wang, N., Manchester, R.~N., Wu, X.~J., \& Wang, H.~X.\ 2005,
  Monthly Notices of the Royal Astronomical Society, 362, 1189



\end{thebibliography}
\end{document}